\documentclass[preprint,showpacs,preprintnumbers,superscriptaddress]{revtex4}
\usepackage[colorlinks,linkcolor=blue,anchorcolor=blue,citecolor=blue]{hyperref}
\usepackage{amsmath}
\usepackage{amssymb}
\usepackage{amsfonts}
\usepackage{subfigure}
\usepackage{graphicx}% Include figure files
\usepackage{dcolumn}% Align table columns on decimal point
\usepackage{bm}% bold math

\newcommand{\reftab}[1]{TABLE \ref{#1}}
\newcommand{\reffig}[1]{FIG. \ref{#1}}
\newcommand{\refeqs}[1]{Eqs. (\ref{#1})}

\begin{document}

\title {Optimizing doping parameters of target to enhance direct-drive implosion}

\author{Guannan Zheng}
\affiliation{Department of Plasma Physics and Fusion Engineering and CAS Key Laboratory of Geospace Environment, University of Science and Technology of China, Hefei, Anhui 230026, People's Republic of China}

\author{Tao Tao}
\email[e-mail:]{tt397396@ustc.edu.cn}
\affiliation{Department of Plasma Physics and Fusion Engineering and CAS Key Laboratory of Geospace Environment, University of Science and Technology of China, Hefei, Anhui 230026, People's Republic of China}

\author{Qing Jia}
\affiliation{Department of Plasma Physics and Fusion Engineering and CAS Key Laboratory of Geospace Environment, University of Science and Technology of China, Hefei, Anhui 230026, People's Republic of China}
	
\author{Rui Yan}
\affiliation{Department of Modern Mechanics, University of Science and Technology of China, Hefei, Anhui 230026, People's Republic of China}
\affiliation{Collaborative Innovation Center of IFSA, Shanghai Jiao Tong University, Shanghai 200240, People's Republic of China}

\author{Jian Zheng}
\email[e-mail:]{jzheng@ustc.edu.cn}
\affiliation{Department of Plasma Physics and Fusion Engineering and CAS Key Laboratory of Geospace Environment, University of Science and Technology of China, Hefei, Anhui 230026, People's Republic of China}
\affiliation{Collaborative Innovation Center of IFSA, Shanghai Jiao Tong University, Shanghai 200240, People's Republic of China}

\begin{abstract}
Direct-drive is an important approach to achieving the ignition of inertial confinement fusion. To enhance implosion performance while keeping the risk of hydrodynamic instability at a low level, we have designed a procedure to optimize the parameters of the target doped with mid- or high-$Z$ atoms. In the procedure, a one-dimensional implosion can be automatically simulated, while its implosion performance and high-dimensional instability are integrally evaluated at the same time. To find the optimal doping parameters, the procedure is performed in the framework of global optimization algorithm, where we have used the particle swarm optimization in the current work. In the optimization, the opacity of mixture materials is quickly obtained by using an interpolation method, showing only a slight difference from the data of TOPS, which is an online doping program of Los Alamos National Laboratory. To test the procedure, optimization has been carried out for the CH ablator in the double cone ignition scheme [Phil. Trans. R. Soc. A. 378.2184 (2020)] by doping with Si and Cl. Both one- and two-dimensional simulations show that doping with either Si or Cl can efficiently mitigate the instability during the acceleration phase and does not result in significant degradation of the peak areal density. The results from one- and two-dimensional simulations qualitatively match with each other, demonstrating the validity of our optimization procedure.
\end{abstract}

\maketitle
	
\section{Introduction}
\label{Introduction}
Since the concept of inertial confinement fusion (ICF)~\cite{nuckolls1972laser} was introduced in the 1970s, it has been studied in laboratories by a large number of researchers. Direct-drive~\cite{craxton2015direct,campbell2017laser} and indirect-drive~\cite{haan2011point,edwards2011experimental,betti2016inertial} are the two main approaches to achieving fusion ignition. Compared with indirect-drive ICF, direct-drive allows the laser energy to be coupled more efficiently with the implosion target. In direct-drive implosions, a capsule filled with frozen deuterium-tritium (DT) fuel is directly irradiated with the symmetrically arranged high-energy laser beams. The target shell is accelerated inward under the action of ablation pressure, and the DT fuel is cmpressed to extreme conditions to achieve ignition and burning, resulting energy gain.
	
However, the ablation front of the target is subject to Rayleigh-Taylor instability (RTI)~\cite{takabe1983self,takabe1985self,betti1996self,goncharov1996self1,goncharov1996self2} when the DT fuel is accelerated under the ablation pressure.The RTI has been the main obstacle to achieving ignition since the beginning of laser fusion studies, because it can amplify the very small seeds either from laser illumination uniformity or from target deficit~\cite{ishizaki1997propagation,ishizaki1998model,igumenshchev2013effects,hu2016understanding} to large amplitude, destroying the integrity of the imploded shell~\cite{radha2005two,hu2009neutron,hu2010two,baumgaertel2014observation}. To deal with these difficulties, many schemes have been proposed. One way to suppress the RTI is to start with the design of laser pulses and targets, including using the entropy-shaping laser pulses~\cite{metzler2003laser,goncharov2003improved,anderson2004laser} to reduce the growth rate of RTI and optimizing the structure of the target such as using foam-buffered target~\cite{watt1998laser,metzler2002laser,hu2018mitigating}, utilizing laminated ablator~\cite{masse2007stabilizing,masse2011observation}, etc. These strategies have been proven useful for enhancing the target performance. In addition, using mid- or high-$Z$ dopants~\cite{hu2012mitigating,fiksel2012experimental} or high-$Z$ thin-layer coatings~\cite{obenschain2002effects,mostovych2008enhanced,karasik2015suppression} can also suppress laser imprinting and mitigate RTI.
	
In experiments, the implosion of targets is a multi-dimensional hydrodynamic process. To achieve ignition, we need to reduce the level of RTI while keeping the implosion at high performance to ensure the integrity of the shell and achieve symmetric compression as much as possible. Although the above-mentioned strategies have been theoretically and experimentally proven effective in suppressing instabilities and enhancing implosions, simulation is still necessary to help us design the targets and laser pulses for the different direct-drive ignition cases, due to the complexity of implosion. In this paper, we integrate the simulation tools and optimization algorithms into one procedure to realize an automatic optimization of target parameters, including the size and doping parameters. In the optimization, we use the one-dimensional (1D) code MULTI~\cite{ramis1988multi,ramis2012multi,ramis2016multi} to simulate the implosions and a semi-analytical model is used to estimate the RTI growth factors. The procedure is performed in the framework of global optimization algorithm to find the optimal target parameters, and we have used the particle swarm optimization (PSO)~\cite{poli2007particle} in the current work. 	A simple method to generate the opacity of mixture materials is developed to take into account the effects of doping on implosion. At last, the multi-dimensional code FLASH~\cite{Fryxell_2000} is used to verify and output the optimal results.
	
To verify the feasibility of our procedure, we have applied it to the double-cone ignition (DCI) scheme~\cite{zhang2020double}. DCI is a novel direct-drive scheme proposed by J. Zhang et al in 2020, consisting of four subtle controllable processes: quasi-isentropic compression, acceleration, head-on collision, and magnetic field-guided fast heating. In this paper, we have doped the polystyrene (CH) ablator of the DCI target with Si or Cl atoms and find that the stability of the doped target is greatly increased while the areal density of the fuel keeps reasonably high, which is verified with our 2D simulations. 
	
This paper is arranged as follows. In Sect. \ref{Optimization procedure}, we focus on the detailed descriptions of each module for the optimization. In Sect. \ref{Optimization of DCI}, we perform a doping optimization for the target of the DCI scheme. Later in this section, the 2D simulations are also carried out with FLASH for the previous 1D optimization parameters, whose results are analyzed and compared with 1D simulations. Finally, a summary is given out in Sect. \ref{Conclusion}.
	
\section{Optimization procedure}
\label{Optimization procedure}
	
\subsection{Overview}
\label{Overview}
	
\begin{figure}
	\centering
	\includegraphics[width=0.96\textwidth]{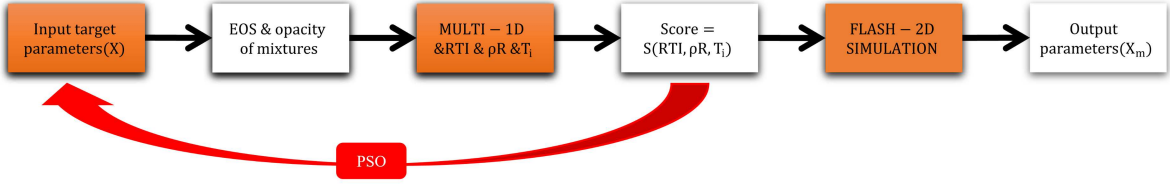}
	\caption{Schematic diagram of the optimization procedure. The 1D simulations are performed in the first four modules within the framework of global optimization algorithm (PSO, GA, etc.) to find the optimal parameters. Before the output, the previous results will be verified by the 2D simulations with FLASH.}
	\label{procedure}
\end{figure}

In this part, we first introduce the procedure we have built to automatically dope the direct-drive target with mid-/high-$Z$ atoms and optimize its parameters. The procedure can be divided into several parts, which are schematically shown in \reffig{procedure}. By inputting the target parameters ${X}$ (e.g. the target size, mid-/high-$Z$ atoms type, and doping ratio, etc.), the procedure will generate a corresponding simulation case, and use the doping module to generate the equation of state (EOS) and opacity data of the target materials. After that, a 1D implosion simulation is carried out and a post-processor is developed to compute the RTI gain and implosion performance (e.g. the peak areal density ${\rho R}$, ion temperature ${T_i}$, etc.) to produce a comprehensive value $Score$. To find the optimal target parameters $X_m$ corresponding to the highest value $Score_m$, the first four modules in \reffig{procedure} will work cyclically in the framework of a global optimization algorithm (e.g. genetic algorithm GA, particle swarm optimization PSO, etc.) until the procedure converges and gives out the 1D optimization results. Before the output, the 1D optimization parameters need to be verified by the 2D simulations with FLASH. In 1D simulations, we have used the radiation hydrodynamic code MULTI with a Lagrangian grid, including both the multigroup radiation transport and $\mathrm{Spitzer-H\ddot{a}rm}$ electron thermal conduction model with a flux limiter of 0.06.
	
\subsection{Opacity of mixtures}
\label{Opacity of mixtures}
	
EOS and opacity are the data related to the physical properties of the material, and they play a crucial role in the simulations. The EOS used for CH and DT are obtained from the MULTI source program and the opacities of C, H, Cl, Si, and DT are from the 40-group ATOMIC opacity tables of TOPS~\cite{abdallah1985tops,magee1995atomic}, which is an online doping program developed in Los Alamos National Laboratory (LANL). In current optimization cases, the doping ratio of mid-$Z$ atoms is low and the opacity, which dominates the radiation transport in ICF, shows a greater influence than EOS in our simulations. Therefore, we have used the EOS data for CH instead of dopants in the optimization, but the opacity is the real data of mixtures.
	
To integrate the doping process into the automatic optimization procedure, we have developed a quick interpolation method to obtain the mixture opacity. This method is inspired by the opacity formula of fully ionized plasmas, which is written as~\cite{zel2002physics,atzeni2004physics}
\begin{equation}
\kappa_{\nu} = 2.78\frac{Z^3\rho^2}{A^2\sqrt{T_e}(h\nu)^3}~\text{cm}^{-1},
\label{ff opacity1}
\end{equation}
where $Z, A, \rho, T_e, h, \text{and}\ \nu$ represent the nuclear charge, mass number of ions, plasma density, electron temperature, Planck's constant, and the frequency of photons respectively. We write this formula as
\begin{equation}
\kappa_{\nu} = K(Z,\nu,T_e)N_eN_i,\label{ff opacity2}
\end{equation}
where $N_e$ and $N_i$ represent the electron and ion density of plasmas. In Eq. (\ref{ff opacity2}), $K(Z,\nu,T_e)$ is the coefficient indicating the ability of an ion to absorb or emit the photons of frequency ${\nu}$ when the ion and electron are scattered with each other. From Eq. (\ref{ff opacity2}), it can be found that in fully ionized plasmas, the opacity is proportional to the product of electron and ion densities. In our doping process, we assume that the ionization properties of an atom and its absorption properties (including the line-absorption (b-b), photoionization (b-f), and inverse bremsstrahlung absorption (f-f)) are also only determined by the temperature and total electron density of the plasmas.

\begin{figure}[ht]
	\centering
	\includegraphics[width=0.49\textwidth]{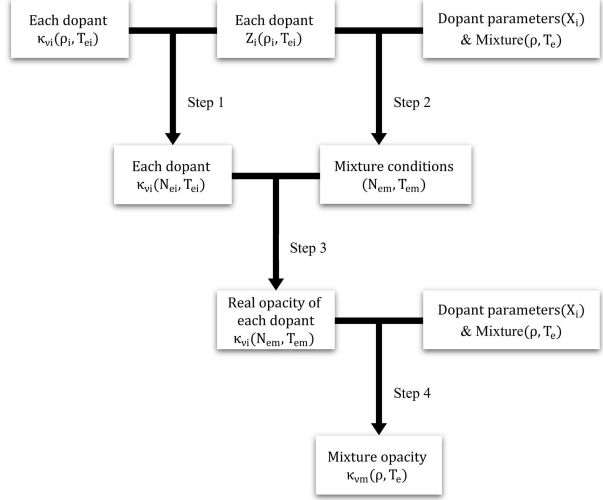}
	\caption{Schematic diagram of the interpolation method to calculate the opacity data of mixture materials.}
	\label{doping}
\end{figure}
	
The doping method is schematically described in \reffig{doping}. In step 1, the opacity table $\kappa_{\nu i}(\rho_i,T_{ei})$ of each dopant, which represents the opacity data in unit $\text{cm}^2/\text{g}$, will be converted into $\kappa_{\nu i}(N_{ei},T_{ei})$ with its ionization table $Z_i(\rho_i,T_{ei})$. Here we have used the subscript $i$ to number the dopants. In step 2, the total electron density of the mixture, $N_{em}(\rho, T_e)$, can also be obtained from the doping parameters $X_i$ by neglecting the interactions of ionization between these dopants. 	When the mixture is in the condition of $(\rho, T_e)$, the electron density and temperature felt by each dopant is $(N_{em}, T_{em}=T_e)$. To get the real opacity of each dopant, we need to interpolate the opacity table $\kappa_{\nu i}(N_{ei},T_{ei})$ to ${(N_{em}, T_{em})}$ in step 3. Finally, in step 4, these dopants' data will be weighted and averaged according to the ratio of the atomic number of each dopant.
	
\begin{figure}
\centering
\includegraphics[width=0.48\textwidth]{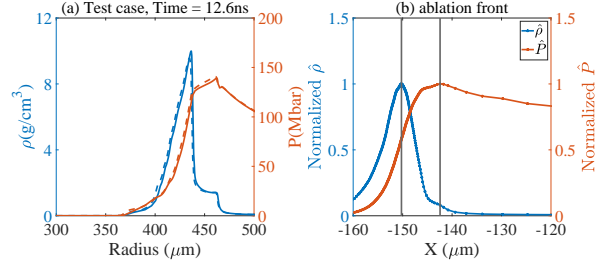}
\caption{(a) Benchmark case at 12.6 ns with a 1D spherical implosion of the DCI target in \reffig{target14}(a), where the solid line uses the mixture opacity from TOPS while the dashed line corresponds to our interpolation method. (b) The ablation front structure of 1D laser-driven slabs, where the area between two grey lines satisfies the condition ${\nabla\rho\cdot\nabla P<0}$. In the unstable region, the normalized density and pressure need to be fitted to \refeqs{density_profile} and (\ref{pressure_profile}) to obtain the ablation parameters such as ${\nu}$, ${L_0}$, ${Fr}$, etc.}
\label{test&ablation}
\end{figure}
	
Although it may be doubtful to ignore the interactions of ionization between the dopants and to assume that the opacity depends only on the total electron density and temperature, these assumptions are shown acceptable in our implosion simulations. We have compared the mixture opacity data obtained by this interpolation method with TOPS, and the simulation results are shown in \reffig{test&ablation}(a), where we have doped the CH ablator of the DCI target with ${2\%}$ Si. In \reffig{test&ablation}(a), the dashed line represents the implosion that uses the opacity from our interpolation method, while the solid line corresponds to TOPS. There is only a slight difference between the two simulations, demonstrating the validity of our doping approach.

\subsection{RTI growth factor}
\label{RTI growth factor}
	
During the implosion, the ablation front accelerated inward is subject to the RTI, whose linear growth rate can be written as a modified Takabe formula~\cite{betti1998growth}:
\begin{equation}
\gamma=\alpha(\nu,F_r)\sqrt{\frac{kg}{1+kL_m}}-\beta(\nu,F_r) kV_a,\label{RTI_formula}
\end{equation}
where ${\alpha}$ and ${\beta}$ are the two coefficients dependent on the dimensionless Froude number ${Fr}$ and effective thermal conductivity power index ${\nu}$. The variables ${g}$, ${L_m}$ and ${V_a}$ represent the acceleration of ablation front, minimum value of density scale length, and ablation velocity, respectively. The density and pressure of the ablation front are shown in \reffig{test&ablation}(b), where we have normalized the profile to its peak value. The RTI occurs when the pressure and density meet the condition ${\nabla\rho\cdot\nabla P<0}$. In \reffig{test&ablation}(b), we have also marked the unstable region (the area between two grey lines). To estimate the linear growth factor of RTI, we need to fit the normalized density and pressure to the differential \refeqs{density_profile} and (\ref{pressure_profile}) satisfied by the steady-state isobaric ablation model~\cite{betti1998growth} to get the parameters in Eq. (\ref{RTI_formula}),
\begin{equation}
L_0\frac{d\hat{\rho}}{dx}=\hat{\rho}^{\nu+1}(\hat{\rho}-1),\label{density_profile}
\end{equation}
\begin{equation}
\frac{1}{\Pi_a^2}\frac{d\hat{P}}{dx}=\frac{1}{\hat{\rho}^2}\frac{d\hat{\rho}}{dx}+\frac{\hat{\rho}}{F_rL_0},\label{pressure_profile}
\end{equation}
where ${\hat{\rho}}$ and ${\hat{P}}$ represent the normalized density and pressure, and ${L_0}$ is the characteristic thickness of the ablation front, satisfying 
$${L_m=L_0\frac{(\nu+1)^{(\nu+1)}}{\nu^\nu}}.$$
In Eq. (\ref{pressure_profile}), the parameter ${\Pi_a}$ is the normalized ablation velocity. The acceleration ${g}$ and ablation velocity ${V_a}$ in Eq. (\ref{RTI_formula}) can be obtained by the relations $g={V_a^2}/{(F_rL_0)}$ and ${V_a=\Pi_a\sqrt{P_a/\rho_a}}$, where ${P_a}$ is the pressure at the location of the peak density ${\rho_a}$. At each moment, a post-processor for MULTI is developed to fit the density and pressure profiles to calculate the RTI growth rate. After time integration ${\int\gamma(t) d t}$, the RTI gain factor in the acceleration phase is obtained.

\newpage

\subsection{Score function}
\label{Score function}
	
\begin{figure}
	\centering
	\includegraphics[width=0.4\textwidth]{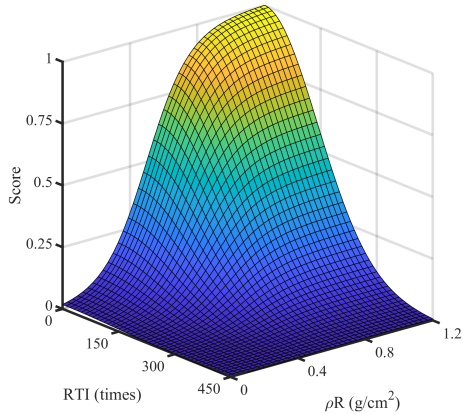}
	\caption{Schematic diagram of the score function, an \textit{inverted bell} surface obtained by multiplying both the normalized Sigmoid and Gaussian functions. The Sigmoid function is used to score the implosion performance (${\rho R}$, ${T_i}$, etc.), while the Gaussian function is for the RTI.}
	\label{score}
\end{figure}

Considering both the 1D implosion performance and multi-dimensional RTI, we score the implosion process with a value that is dependent on the RTI growth factor and peak areal density. In the current work, the score function is an \textit{inverted bell} surface obtained by multiplying both the normalized Sigmoid and Gaussian functions, as shown in \reffig{score}. The Sigmoid function is used to score the compressed areal density of imploded fuel, while the Gaussian function is for the RTI. Once the previous modules are built, an implicit function mapping from the target and laser parameters to the comprehensive score of implosion can be realized. Note that the form of the score function is important to the optimization results, and its current form is rather simple. If necessary, it would be revised based on the results of high-dimensional simulations and experiments. Our goal is to find the target parameters with the highest score. As a swarm intelligence algorithm, PSO is a heuristic global optimization algorithm that uses multiple particles to find the optimal solutions in parameter space, thus it is naturally suitable for large-scale parallel computing, greatly increasing the optimization efficiency.

\section{Optimization of DCI}
\label{Optimization of DCI}

\subsection{DCI scheme}
\label{DCI scheme}
	
\begin{figure}
	\centering
	\includegraphics[width=0.4\textwidth]{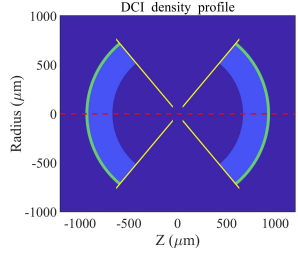}
	\caption{A 2D schematic for DCI density profile, which is cylindrically symmetric about the red dashed line. In the scheme, the overall radius of the DCI target is $\mathrm{944\ \mu m}$.}
	\label{dci2020}
\end{figure}

With the procedure mentioned above, we perform a target optimization for the case of quasi-isentropic laser pulse in the article by J. Zhang et al~\cite{zhang2020double}. In the DCI scheme, the compression and acceleration of DT are carried out inside two head-on cones, whose angles are ${100^{\circ}}$, corresponding to the solid angle ${\Omega=2.24}$, as shown in \reffig{dci2020}. Compared with the scheme of hot spot, DCI theoretically requires only ${4.48/4\pi}$ of the total energy needed by spherically symmetric implosion to achieve the same implosion performance, significantly reducing the demand for driving laser energy. 	Next, the accelerated fuel will be ejected from each cone and collide in the central area of about $\mathrm{100\ \mu m}$, further increasing the density and temperature of compressed DT. Finally, the MeV electrons guided by a strong magnetic field will further heat the fuel, bringing it to the ignition condition.

\begin{figure}
\centering
\includegraphics[width=0.48\textwidth]{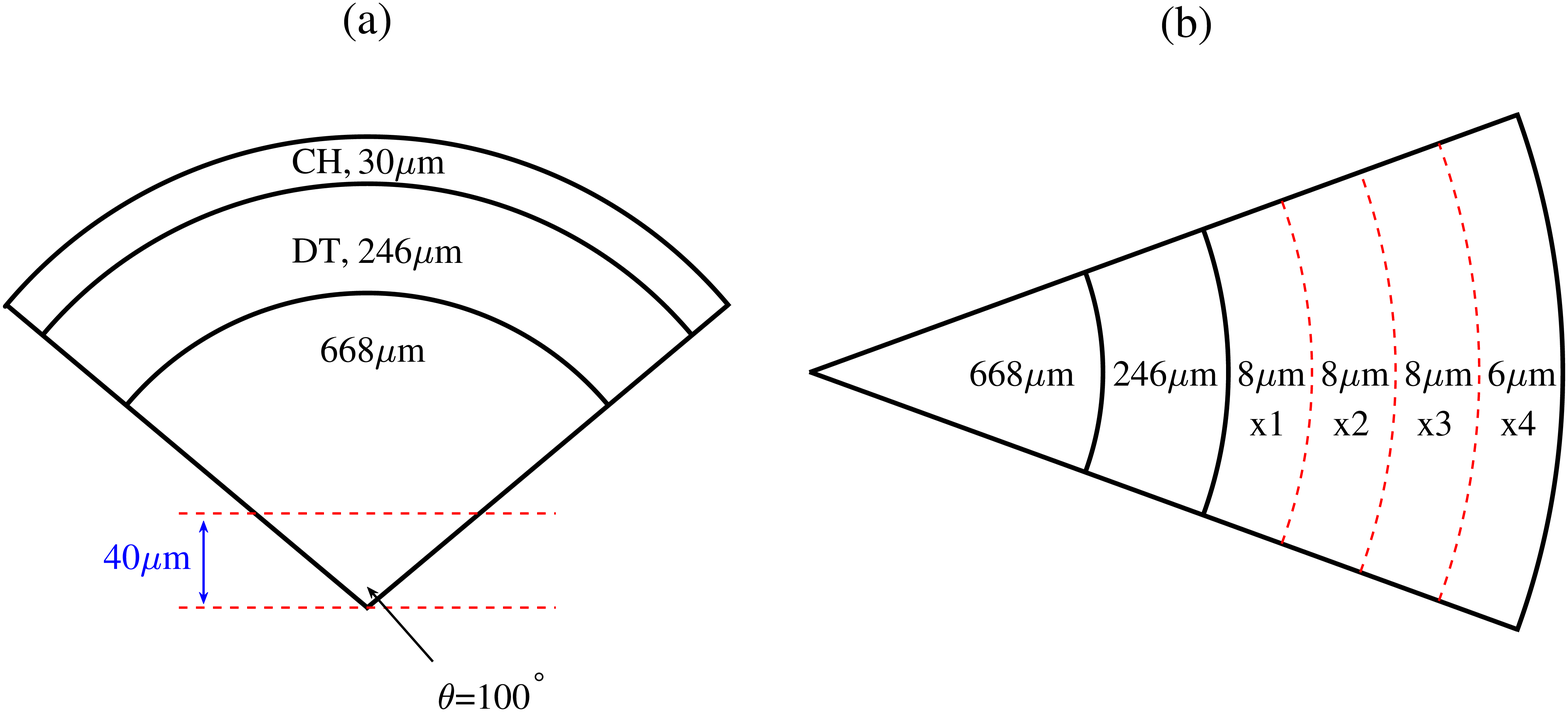}
\caption{(a) Schematic diagram of the DCI target, where the densities of CH and DT are ${1.1}~\text{g/cm}^3$ and $0.25~\text{g/cm}^3$ respectively. (b) Schematic diagram of the target in layered-doping optimization. The $30~\mu\text{m}$ ablator is divided into 4 layers, whose thickness are $8~\mu\text{m}$, $8~\mu\text{m}$, $8~\mu\text{m}$, and $6~\mu\text{m}$, respectively. In the optimization, the thickness of each layer is kept constant. The variables x1, x2, x3, and x4 represent the abundances (${\%}$) of dopants (Si or Cl) in corresponding layers.}
\label{target14}
\end{figure}
	
The target of DCI is composed of $246~\mu\text{m}$ DT ice with the $668~\mu\text{m}$ inner radius and $30~\mu\text{m}$ CH ablator, as shown in \reffig{target14}(a). In the figure, we have also marked the position of cone tips with a red dashed line, about $40~\mu\text{m}$ away from the center of the plastic shell. 	The asymmetry of DCI geometry and the existence of expanding gold plasma may lead to an increased risk of the RTI. We choose Si as a dopant to optimize the implosion with the guidance of previous studies~\cite{hu2012mitigating,fiksel2012experimental}. In DCI experiments performed on the SG-IIU facility, $\mathrm{C_{16}H_{14}C_{12}}$ has been used as the ablation material instead of $\mathrm{C_{8}H_{8}}$. Due to the severe radiation preheating caused by the high proportion of Cl atoms in the ablator, MULTI-1D simulations show that the compressed areal density is severely reduced compared to that of the plastic shell. Therefore, in the optimization, we also try using a low proportion of Cl to improve the implosion performance. 	In the current work, we optimize the CH ablator with the single-layer and multi-layer (4-layer, currently as a simple test case) doping of Si and Cl. In the multi-layer optimization, the thickness of each layer remains unchanged and the layered structure is shown in \reffig{target14}(b), where x1, x2, x3, and x4 represent the dopant atomic (Cl or Si) percentage in corresponding layers.

\begin{table}
	\caption{1D-optimization results}\label{table1}
	\centering
	\begin{tabular}{c c c c c c c p{2cm}<{\centering} p{2cm}<{\centering}} \hline\hline
		Case  & Dopant & Layer & x1($\%$)&x2($\%$)&x3($\%$)&x4($\%$)&RTI gain &$\rho R~\mathrm{(g/cm^2)}$\\
		\hline
		CH    & --     & 1     &0.0&0.0&0.0&0.0&$\gg 100$&0.85\\
		
		Cl-1  & Cl     & 1     &0.6&0.6&0.6&0.6&26&0.77\\
		
		Cl-4  & Cl     & 4     &0.0&2.0&0.0&0.6&22&0.81\\
		
		Si-4  & Si     & 4     &0.0&2.4&0.3&1.4&11&0.85\\
		
		Si-4  & Si     & 4     &0.0&2.6&0.4&1.4&11&0.85\\
		
		Si-4  & Si     & 4     &0.0&2.7&0.3&1.4&12&0.85\\
		\hline\hline
	\end{tabular}
\end{table}
	
\subsection{1D-optimization}
\label{1D-optimization}
\begin{figure}[ht]
\centering
\subfigure{
\includegraphics[scale=0.16]{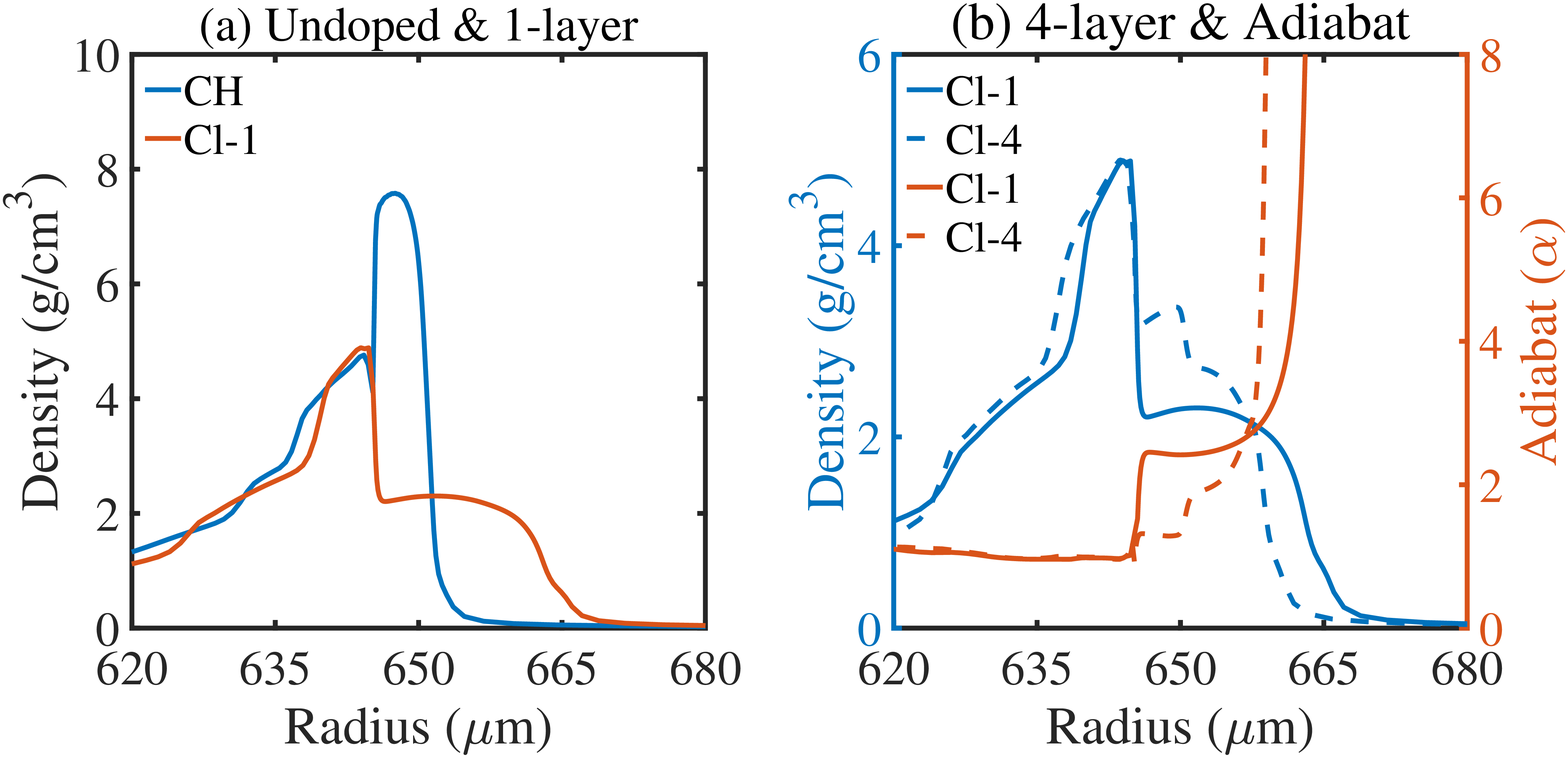}
}
\subfigure{
\includegraphics[scale=0.16]{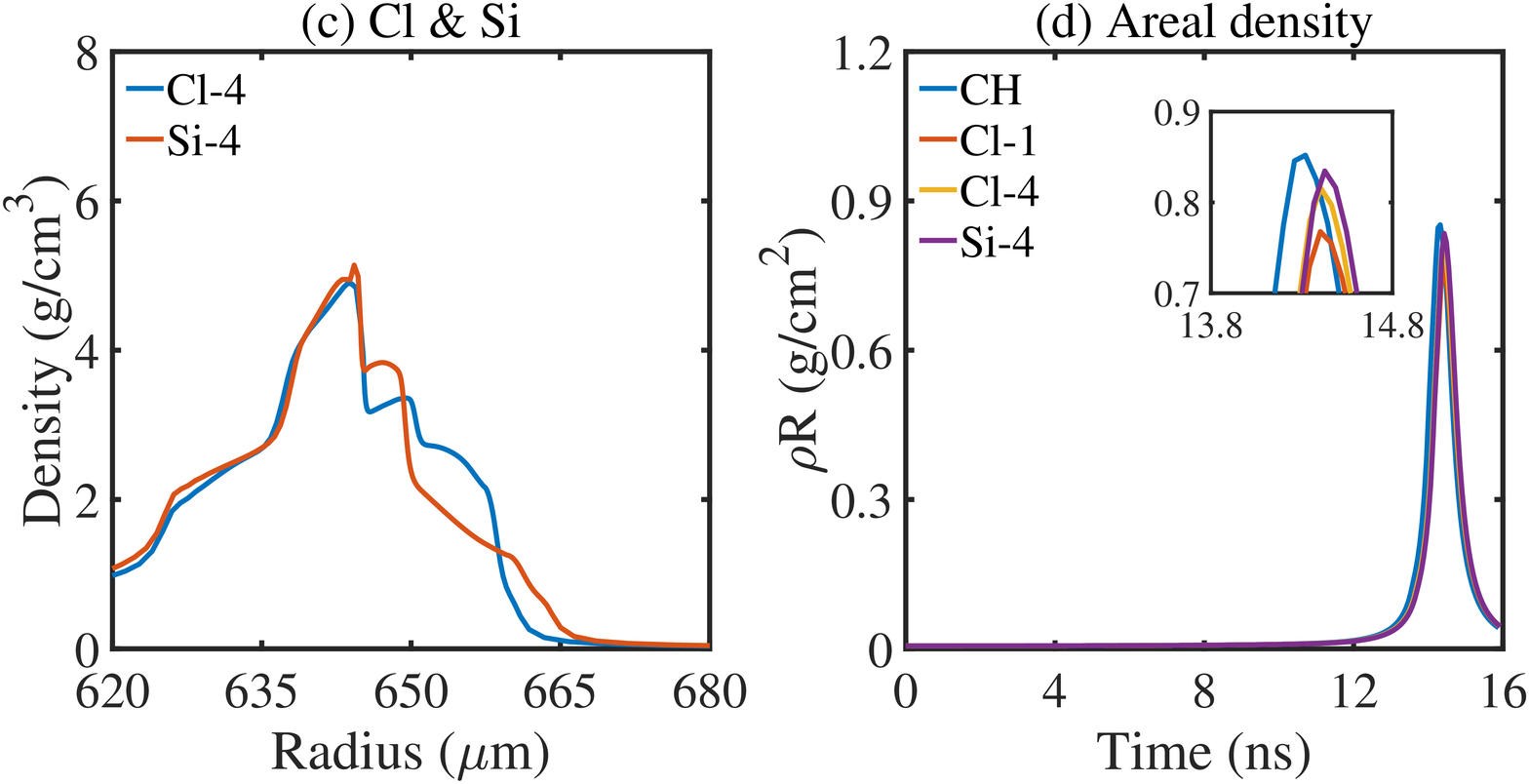}
}
\caption{(a) Density comparison of the undoped case CH and the single-layer doping case Cl-1. (b) Schematic diagram of the density and adiabat (${\alpha}$) profile of the case Cl-1 and Cl-4. (c) Density comparison of the case Cl-4 and Si-4. The previous three diagrams are at 10.5 ns. (d) Time evolution of the areal density of the case CH, Cl-1, Cl-4, and Si-4.}
\label{multi1d-10.5ns}
\end{figure}
	
\reftab{table1} shows the 1D-optimization results, where the cases Cl-1 and Cl-4 represent the single-layer and four-layer doping with Cl, respectively. For comparison, the undoped scheme, the case CH, is also shown in the table. The RTI gain factor of CH estimated by the 1D-optimization program is ${\gg 100}$, while its areal density reaches $\mathrm{0.85~ g/cm^2}$. Due to the differences such as the EOS and opacity data of materials, the peak areal density in our simulations is slightly higher than $\mathrm{0.82~g/cm^2}$ in the article~\cite{zhang2020double}. By doping CH ablator with Cl, the gain factor of RTI is significantly reduced (${\textless 30}$). The case Cl-4 has a comparable RTI gain factor to Cl-1, but the areal density is increased from $\mathrm{0.77\ to\ 0.81\ g/cm^2}$. \reffig{multi1d-10.5ns}(a) has compared the density profile of the case CH with Cl-1 at 10.5 ns. In the figure, the density scale length of the ablation front is significantly increased by doping with Cl atoms. \reffig{multi1d-10.5ns}(b) compares the density and adiabat factor (${\alpha}$) of the case Cl-1 with Cl-4, where the adiabat factor represents the entropy increase of materials. In the 4-layer doping scheme Cl-4, the high doping ratio (${\sim 2\%}$) of x2 significantly reduces the entropy increase in the x1 layer, because x2 absorbs most of the radiation energy from x3 and x4. The reduction of entropy results in a relatively high density of the ablator and fuel for compression, increasing the peak areal density. In the x3 and x4 layers of Cl-4, the temperate doping ratio maintains the ablation front through radiation preheating to have a comparable RTI gain factor to the case Cl-1.

As for the dopant Si, we have performed the four-layer doping for the DCI target and \reftab{table1} shows the three independent doping cases. From the table, it can be found that the results of the three optimization cases Si-4 are almost the same, indicating the convergence of our optimization procedure. 	Compared with Cl-4, the case Si-4 has a lower growth factor of the RTI while maintaining a higher areal density ($\mathrm{\sim 0.85\ g/cm^2}$). Furthermore, the doping features of Cl-4 and Si-4 are similar, i.e., the x2 and x4 layers are doped with higher fractions, with x2 being the most abundant. From the density comparison of Cl-4 and Si-4 in \reffig{multi1d-10.5ns}(c), the density scale length at the ablation front of the case Si-4 is larger, while its entropy increase is also lower at the same time. Note that the doping leads to a decrease in the density of ablator due to the radiation preheating of mid-$Z$ atoms, as shown in \reffig{multi1d-10.5ns}(a), (b), and (c), which could lead to an occurrence of RTI at the DT-ablator interface. 	\reffig{multi1d-10.5ns}(d) shows the areal density evolution of the four cases (CH, Cl-1, Cl-4, Si-4), where the simulation parameters of Si-4 are the average of three optimization results, i.e., $\mathrm{x1:x2:x3:x4=0.0:2.53:0.33:1.4\ (\%)}$. Among the three doping schemes, the case Si-4 has the best optimization effects, whose areal density ($\mathrm{\sim 0.85\ g/cm^2}$) is only slightly reduced compared to CH.
	
\subsection{2D-simulation}
\label{2D-simulation}
To verify the doping results, we perform 2D simulations for the case CH, Cl-1, Cl-4, and Si-4. The program used in simulations is the radiation hydrodynamic code FLASH~\cite{Fryxell_2000}, developed by the University of Chicago. The 2D cylindrical grid of FLASH has been used in the simulations, i.e., the program only resolves R and Z directions.
	
\begin{figure}
	\centering
	\subfigure{
		\includegraphics[width=0.48\textwidth]{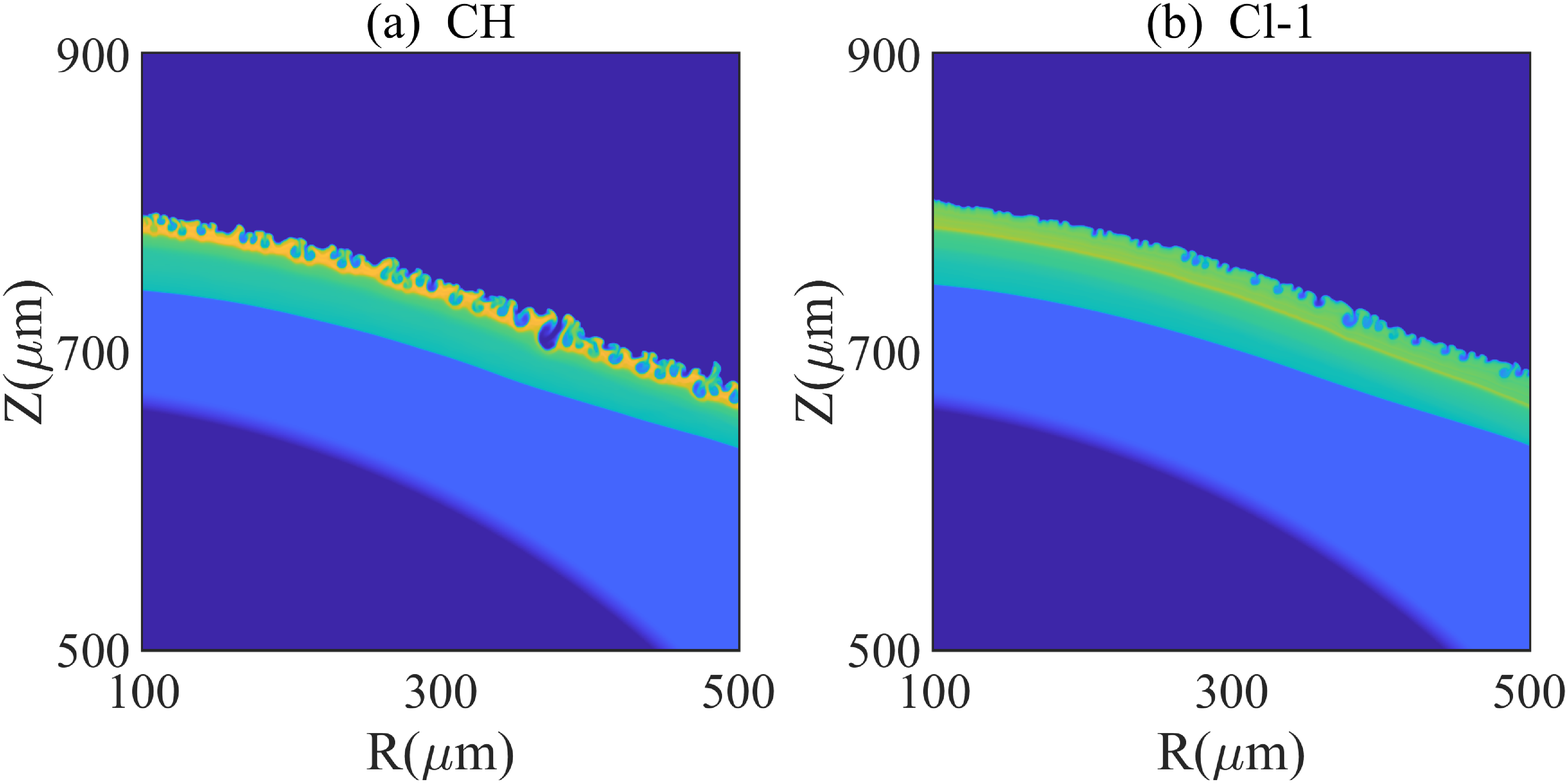}
	}\\
	\subfigure{
		\includegraphics[width=0.48\textwidth]{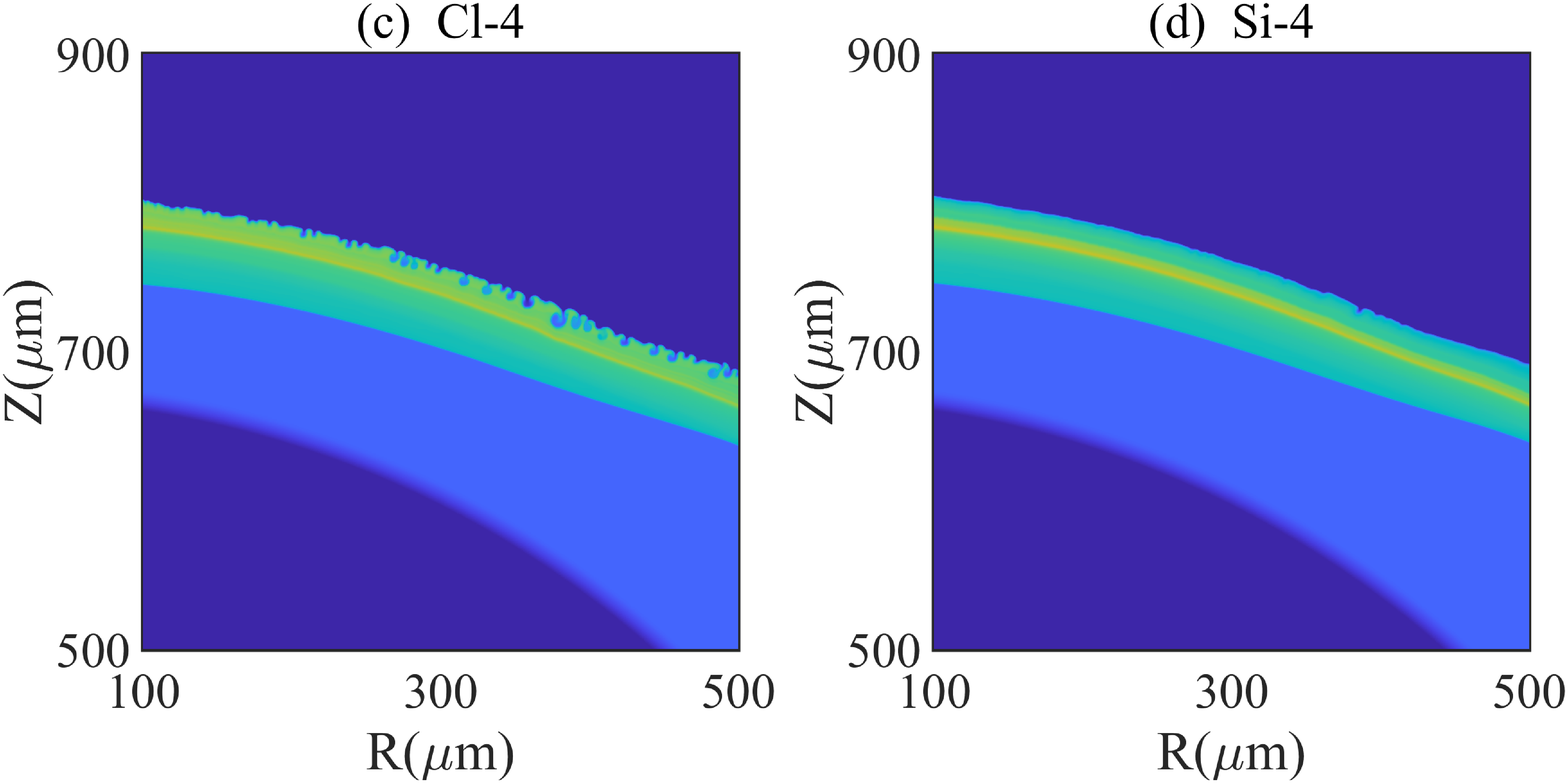}
	}
	\caption{Schematic diagram of the density of (a) undoped case CH, (b) single-layer doping case Cl-1, (c) 4-layer doping case Cl-4, and (d) case Si-4 at 7 ns.}
	\label{flash2d-7ns}
\end{figure}

Firstly, considering the case of there being initial perturbations on the DCI target surface, we have performed high-resolution ($\mathrm{0.5\ \mu m}$) numerical simulations for the four doping schemes, and the simulation results are shown in \reffig{flash2d-7ns}. In the pre-compression ($\mathrm{Time=7\ ns}$), \reffig{flash2d-7ns}(a) shows that the target perturbation of the undoped case CH has developed seriously and evolved to a nonlinear stage. The perturbation developments of the cases Cl-1 and Cl-4 are comparable, while they both are greatly suppressed compared with those of the case CH, as shown in \reffig{flash2d-7ns}(b) and (c). Compared with Cl, the case Si-4 presents a more efficient suppression to the RTI and its perturbation hasn't developed apparently up to 7 ns, as shown in \reffig{flash2d-7ns}(d). From the development of RTI, the 2D results in the pre-compression stage qualitatively verify the conclusions of the 1D-optimization procedure, showing that our 1D-optimization method is effective for suppressing the RTI of DCI.
	
\begin{figure}
	\centering
	\subfigure{
		\includegraphics[width=0.48\textwidth]{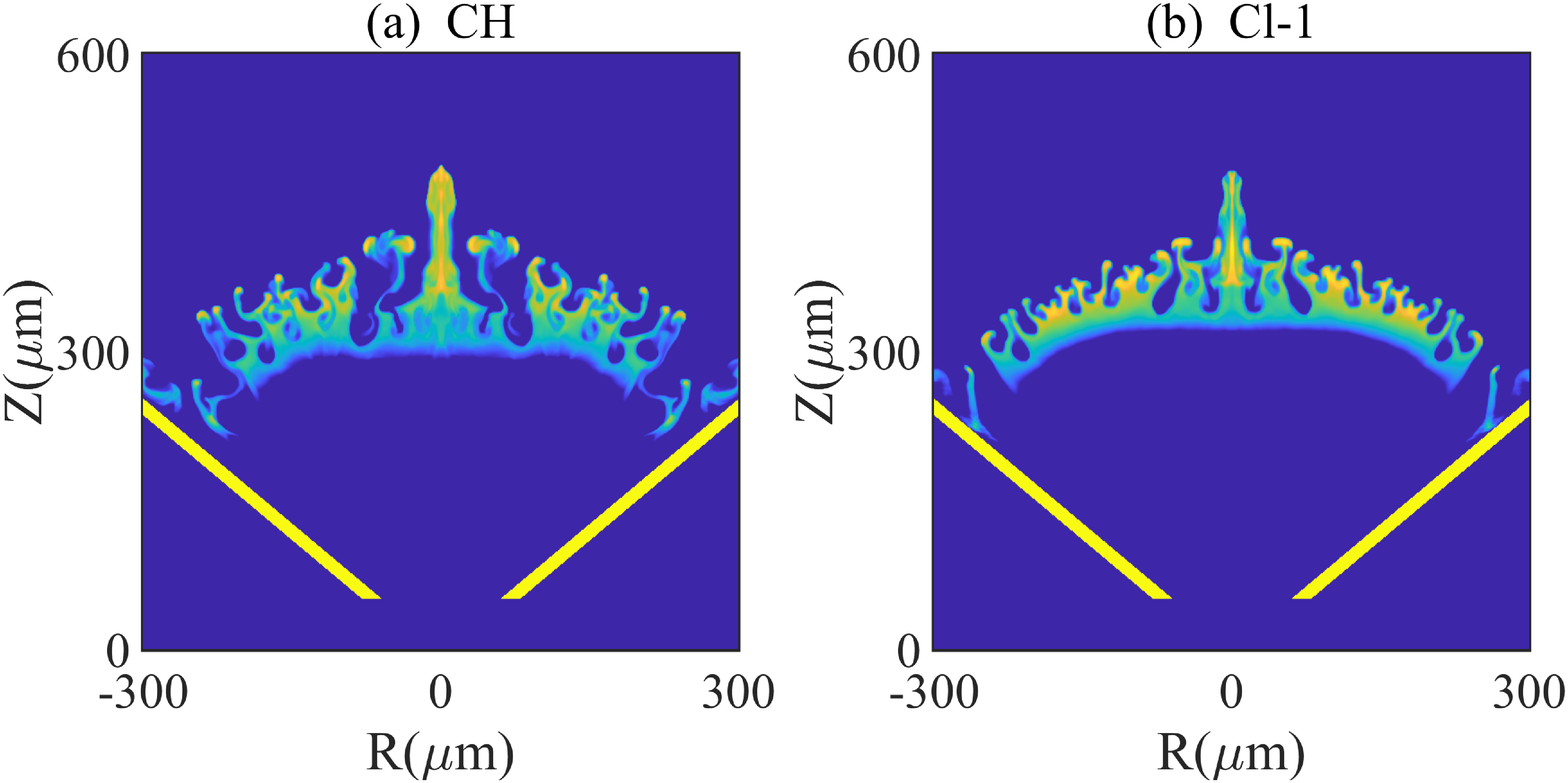}
	}\\
	\subfigure{
		\includegraphics[width=0.48\textwidth]{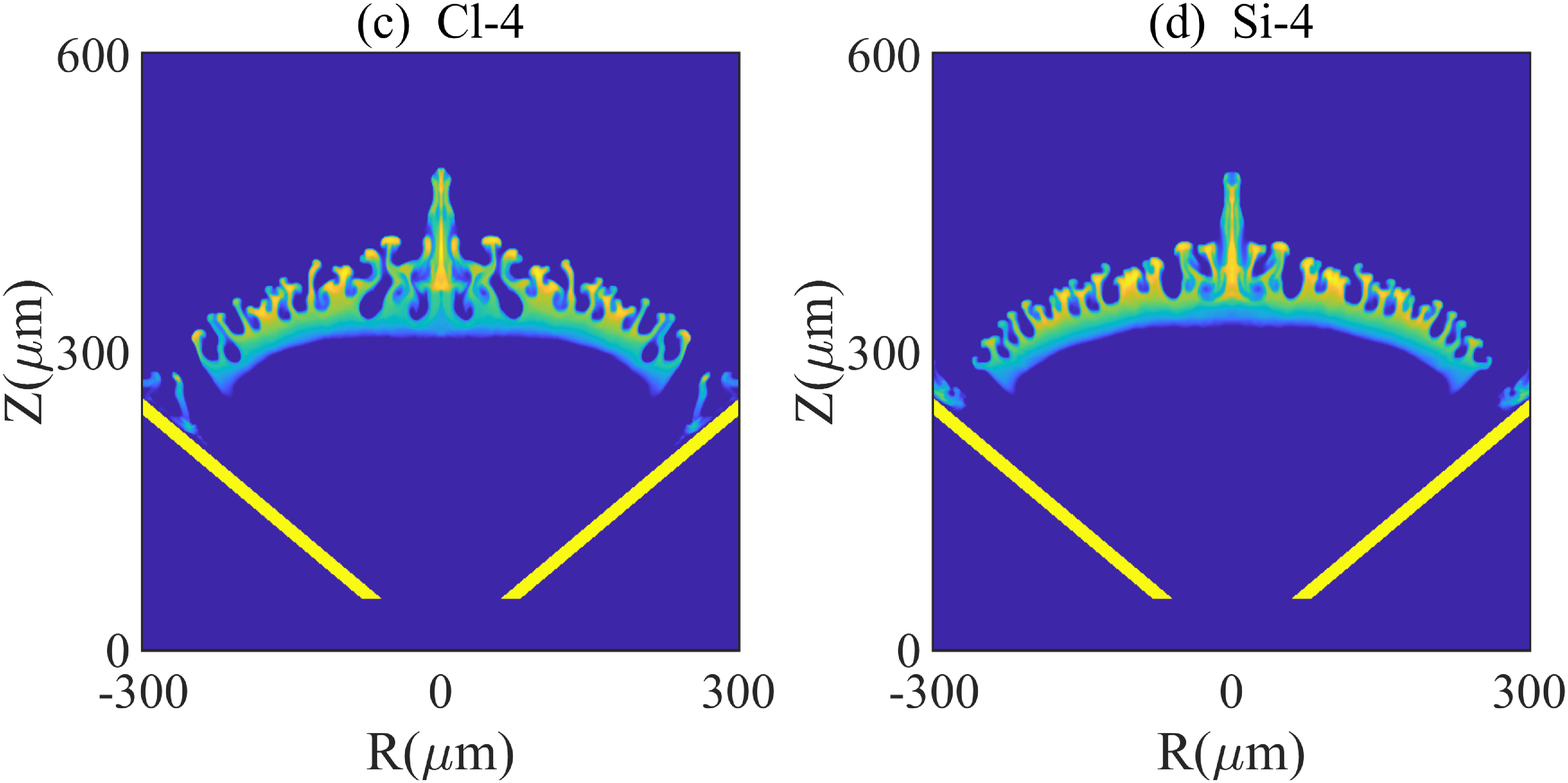}
	}
	\caption{Schematic diagram of the density of the case (a) CH, (b) Cl-1, (c) Cl-4, and (d) Si-4 at 13 ns.}
	\label{flash2d-13ns}
\end{figure}

\begin{figure}
\centering
\includegraphics[width=0.4\textwidth]{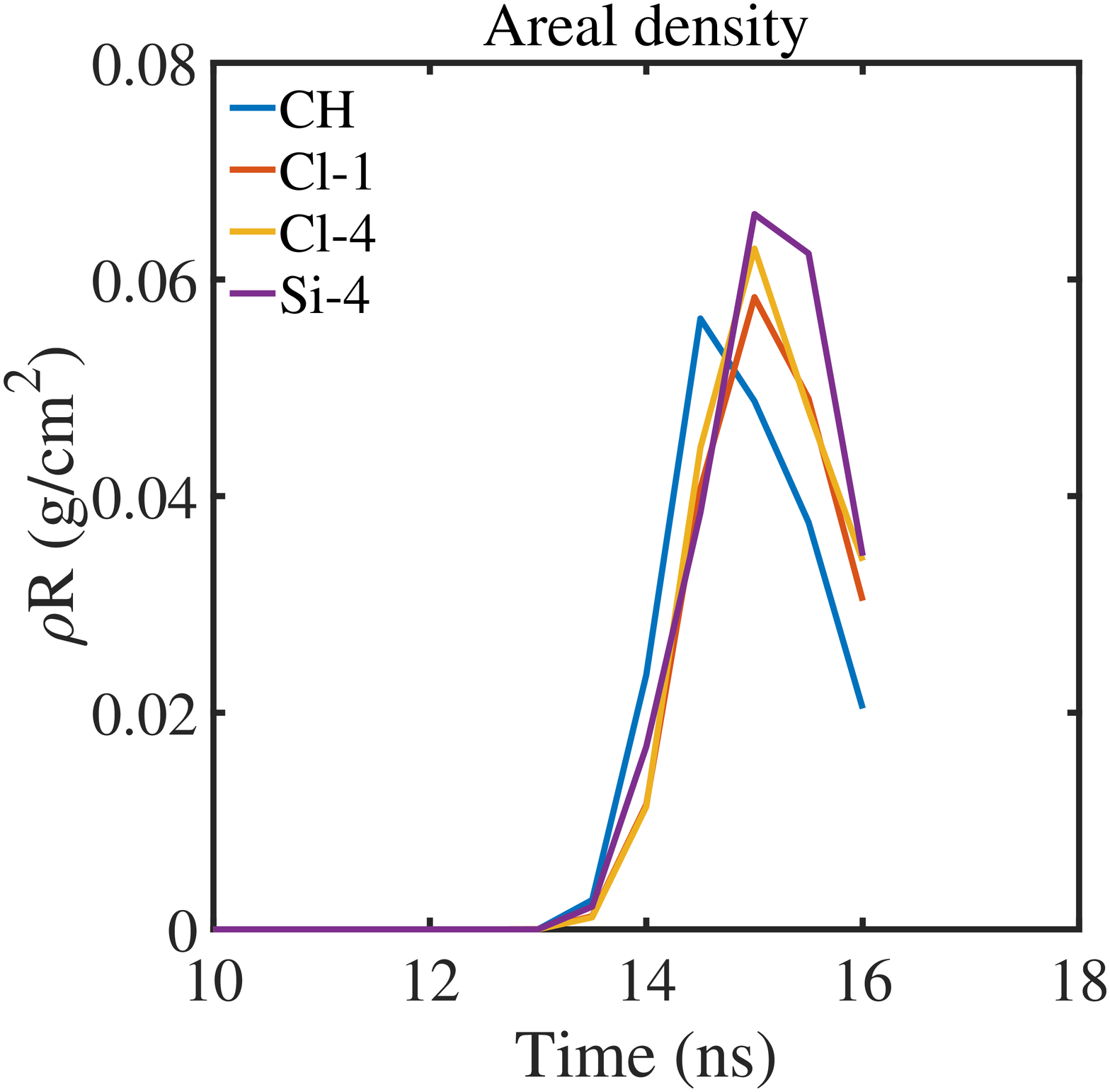}
\caption{Areal density evolution of the case CH, Cl-1, Cl-4, and Si-4.}
\label{flash2d-areal}
\end{figure}

Considering the final areal density of the doping cases, we have also performed the whole 2D simulations with a lower resolution ($\mathrm{1\ \mu m}$). In previous simulations, the initial seeds have developed seriously in the early stage of DCI. So in the whole simulations, we firstly smooth the DCI target surface to reduce the influence of initial grid perturbations. Shown in \reffig{flash2d-13ns} is the density distribution of our 2D simulations during the acceleration phase ($\mathrm{Time=13\ ns}$). It can be found that although the target is moderately smoothed at the initial moment, the RTI would still develop rapidly in the acceleration stage. However, the 2D results from \reffig{flash2d-13ns} still qualitatively match the 1D simulations. The RTI development of the case Si-4, which is shown in \reffig{flash2d-13ns}(d), is weaker than that of the case Cl-1 (\reffig{flash2d-13ns}(b)) and Cl-4 (\reffig{flash2d-13ns}(c)), whose RTIs are comparable. \reffig{flash2d-areal} shows the areal density evolution of DT fuel as the target material passes through the cone tips, where we have averaged the areal density with respect to the theta direction. Due to the severe development of RTI, when the DT fuel reaches the cone tips, its areal density is lower than 1D spherically symmetric implosions. The qualitative conclusions from the areal densities of the cases Cl-1, Cl-4, and Si-4 are also consistent with 1D simulations (\reffig{multi1d-10.5ns}(d)). The compressed areal density of the case CH is the lowest in 2D simulations, due to the severe damage to the target shell integrity.
	
\section{Conclusion}
\label{Conclusion}
	
We propose a procedure to quickly optimize the target design to mitigate the Rayleigh-Taylor instability in the double-cone ignition scheme of direct-drive laser fusion. The RTI gain and areal density are the two key goal parameters in the optimization. A score function that is dependent on the RTI gain and areal density of imploded fuel is thus proposed to evaluate the implosion performance. In the article, we concentrate on the optimization through doping. A simple method is developed to generate the opacity data of the mixture of interests. The opacity data from our method are in excellent agreement with those from LANL's TOPS. The RTI gains and areal density are obtained with the aid of one-dimensional hydrodynamic simulations and Betti's fitting method. The optimal doping parameters are thus quickly found, and the mitigation effects are verified with two-dimensional simulations. It should be pointed out that our optimization method is not only suitable for the quick design of target doping parameters but also for comprehensive optimization of laser pulses and target size, etc.
	
\section*{ACKNOWLEDGMENTS}
	
This work is supported by the Strategic Priority Research Program of the Chinese Academy of Sciences, Grant Nos. XDA25010200, and the Fundamental Research Funds for the Central Universities.
	
\bibliography{dci.bib}

\begin{thebibliography}{43}
\expandafter\ifx\csname natexlab\endcsname\relax\def\natexlab#1{#1}\fi
\expandafter\ifx\csname bibnamefont\endcsname\relax
  \def\bibnamefont#1{#1}\fi
\expandafter\ifx\csname bibfnamefont\endcsname\relax
  \def\bibfnamefont#1{#1}\fi
\expandafter\ifx\csname citenamefont\endcsname\relax
  \def\citenamefont#1{#1}\fi
\expandafter\ifx\csname url\endcsname\relax
  \def\url#1{\texttt{#1}}\fi
\expandafter\ifx\csname urlprefix\endcsname\relax\def\urlprefix{URL }\fi
\providecommand{\bibinfo}[2]{#2}
\providecommand{\eprint}[2][]{\url{#2}}

\bibitem[{\citenamefont{Nuckolls et~al.}(1972)\citenamefont{Nuckolls, Wood,
  Thiessen, and Zimmerman}}]{nuckolls1972laser}
\bibinfo{author}{\bibfnamefont{J.}~\bibnamefont{Nuckolls}},
  \bibinfo{author}{\bibfnamefont{L.}~\bibnamefont{Wood}},
  \bibinfo{author}{\bibfnamefont{A.}~\bibnamefont{Thiessen}}, \bibnamefont{and}
  \bibinfo{author}{\bibfnamefont{G.}~\bibnamefont{Zimmerman}},
  \bibinfo{journal}{Nature} \textbf{\bibinfo{volume}{239}},
  \bibinfo{pages}{139} (\bibinfo{year}{1972}), ISSN \bibinfo{issn}{1476-4687}.

\bibitem[{\citenamefont{Craxton et~al.}(2015)\citenamefont{Craxton, Anderson,
  Boehly, Goncharov, Harding, Knauer, McCrory, McKenty, Meyerhofer, Myatt
  et~al.}}]{craxton2015direct}
\bibinfo{author}{\bibfnamefont{R.~S.} \bibnamefont{Craxton}},
  \bibinfo{author}{\bibfnamefont{K.~S.} \bibnamefont{Anderson}},
  \bibinfo{author}{\bibfnamefont{T.~R.} \bibnamefont{Boehly}},
  \bibinfo{author}{\bibfnamefont{V.~N.} \bibnamefont{Goncharov}},
  \bibinfo{author}{\bibfnamefont{D.~R.} \bibnamefont{Harding}},
  \bibinfo{author}{\bibfnamefont{J.~P.} \bibnamefont{Knauer}},
  \bibinfo{author}{\bibfnamefont{R.~L.} \bibnamefont{McCrory}},
  \bibinfo{author}{\bibfnamefont{P.~W.} \bibnamefont{McKenty}},
  \bibinfo{author}{\bibfnamefont{D.~D.} \bibnamefont{Meyerhofer}},
  \bibinfo{author}{\bibfnamefont{J.~F.} \bibnamefont{Myatt}},
  \bibnamefont{et~al.}, \bibinfo{journal}{Physics of Plasmas}
  \textbf{\bibinfo{volume}{22}}, \bibinfo{pages}{110501}
  (\bibinfo{year}{2015}).

\bibitem[{\citenamefont{Campbell et~al.}(2017)\citenamefont{Campbell,
  Goncharov, Sangster, Regan, Radha, Betti, Myatt, Froula, Rosenberg,
  Igumenshchev et~al.}}]{campbell2017laser}
\bibinfo{author}{\bibfnamefont{E.}~\bibnamefont{Campbell}},
  \bibinfo{author}{\bibfnamefont{V.}~\bibnamefont{Goncharov}},
  \bibinfo{author}{\bibfnamefont{T.}~\bibnamefont{Sangster}},
  \bibinfo{author}{\bibfnamefont{S.}~\bibnamefont{Regan}},
  \bibinfo{author}{\bibfnamefont{P.}~\bibnamefont{Radha}},
  \bibinfo{author}{\bibfnamefont{R.}~\bibnamefont{Betti}},
  \bibinfo{author}{\bibfnamefont{J.}~\bibnamefont{Myatt}},
  \bibinfo{author}{\bibfnamefont{D.}~\bibnamefont{Froula}},
  \bibinfo{author}{\bibfnamefont{M.}~\bibnamefont{Rosenberg}},
  \bibinfo{author}{\bibfnamefont{I.}~\bibnamefont{Igumenshchev}},
  \bibnamefont{et~al.}, \bibinfo{journal}{Matter and Radiation at Extremes}
  \textbf{\bibinfo{volume}{2}}, \bibinfo{pages}{37} (\bibinfo{year}{2017}).

\bibitem[{\citenamefont{Haan et~al.}(2011)\citenamefont{Haan, Lindl, Callahan,
  Clark, Salmonson, Hammel, Atherton, Cook, Edwards, Glenzer
  et~al.}}]{haan2011point}
\bibinfo{author}{\bibfnamefont{S.~W.} \bibnamefont{Haan}},
  \bibinfo{author}{\bibfnamefont{J.~D.} \bibnamefont{Lindl}},
  \bibinfo{author}{\bibfnamefont{D.~A.} \bibnamefont{Callahan}},
  \bibinfo{author}{\bibfnamefont{D.~S.} \bibnamefont{Clark}},
  \bibinfo{author}{\bibfnamefont{J.~D.} \bibnamefont{Salmonson}},
  \bibinfo{author}{\bibfnamefont{B.~A.} \bibnamefont{Hammel}},
  \bibinfo{author}{\bibfnamefont{L.~J.} \bibnamefont{Atherton}},
  \bibinfo{author}{\bibfnamefont{R.~C.} \bibnamefont{Cook}},
  \bibinfo{author}{\bibfnamefont{M.~J.} \bibnamefont{Edwards}},
  \bibinfo{author}{\bibfnamefont{S.}~\bibnamefont{Glenzer}},
  \bibnamefont{et~al.}, \bibinfo{journal}{Physics of Plasmas}
  \textbf{\bibinfo{volume}{18}}, \bibinfo{pages}{051001}
  (\bibinfo{year}{2011}).

\bibitem[{\citenamefont{Edwards et~al.}(2011)\citenamefont{Edwards, Lindl,
  Spears, Weber, Atherton, Bleuel, Bradley, Callahan, Cerjan, Clark
  et~al.}}]{edwards2011experimental}
\bibinfo{author}{\bibfnamefont{M.~J.} \bibnamefont{Edwards}},
  \bibinfo{author}{\bibfnamefont{J.~D.} \bibnamefont{Lindl}},
  \bibinfo{author}{\bibfnamefont{B.~K.} \bibnamefont{Spears}},
  \bibinfo{author}{\bibfnamefont{S.~V.} \bibnamefont{Weber}},
  \bibinfo{author}{\bibfnamefont{L.~J.} \bibnamefont{Atherton}},
  \bibinfo{author}{\bibfnamefont{D.~L.} \bibnamefont{Bleuel}},
  \bibinfo{author}{\bibfnamefont{D.~K.} \bibnamefont{Bradley}},
  \bibinfo{author}{\bibfnamefont{D.~A.} \bibnamefont{Callahan}},
  \bibinfo{author}{\bibfnamefont{C.~J.} \bibnamefont{Cerjan}},
  \bibinfo{author}{\bibfnamefont{D.}~\bibnamefont{Clark}},
  \bibnamefont{et~al.}, \bibinfo{journal}{Physics of Plasmas}
  \textbf{\bibinfo{volume}{18}}, \bibinfo{pages}{051003}
  (\bibinfo{year}{2011}).

\bibitem[{\citenamefont{Betti and Hurricane}(2016)}]{betti2016inertial}
\bibinfo{author}{\bibfnamefont{R.}~\bibnamefont{Betti}} \bibnamefont{and}
  \bibinfo{author}{\bibfnamefont{O.~A.} \bibnamefont{Hurricane}},
  \bibinfo{journal}{Nature Physics} \textbf{\bibinfo{volume}{12}},
  \bibinfo{pages}{435} (\bibinfo{year}{2016}), ISSN \bibinfo{issn}{1745-2481}.

\bibitem[{\citenamefont{Takabe et~al.}(1983)\citenamefont{Takabe, Montierth,
  and Morse}}]{takabe1983self}
\bibinfo{author}{\bibfnamefont{H.}~\bibnamefont{Takabe}},
  \bibinfo{author}{\bibfnamefont{L.}~\bibnamefont{Montierth}},
  \bibnamefont{and} \bibinfo{author}{\bibfnamefont{R.~L.} \bibnamefont{Morse}},
  \bibinfo{journal}{Physics of Fluids} \textbf{\bibinfo{volume}{26}},
  \bibinfo{pages}{2299} (\bibinfo{year}{1983}).

\bibitem[{\citenamefont{Takabe et~al.}(1985)\citenamefont{Takabe, Mima,
  Montierth, and Morse}}]{takabe1985self}
\bibinfo{author}{\bibfnamefont{H.}~\bibnamefont{Takabe}},
  \bibinfo{author}{\bibfnamefont{K.}~\bibnamefont{Mima}},
  \bibinfo{author}{\bibfnamefont{L.}~\bibnamefont{Montierth}},
  \bibnamefont{and} \bibinfo{author}{\bibfnamefont{R.~L.} \bibnamefont{Morse}},
  \bibinfo{journal}{Physics of Fluids} \textbf{\bibinfo{volume}{28}},
  \bibinfo{pages}{3676} (\bibinfo{year}{1985}).

\bibitem[{\citenamefont{Betti et~al.}(1996)\citenamefont{Betti, Goncharov,
  McCrory, Sorotokin, and Verdon}}]{betti1996self}
\bibinfo{author}{\bibfnamefont{R.}~\bibnamefont{Betti}},
  \bibinfo{author}{\bibfnamefont{V.~N.} \bibnamefont{Goncharov}},
  \bibinfo{author}{\bibfnamefont{R.~L.} \bibnamefont{McCrory}},
  \bibinfo{author}{\bibfnamefont{P.}~\bibnamefont{Sorotokin}},
  \bibnamefont{and} \bibinfo{author}{\bibfnamefont{C.~P.}
  \bibnamefont{Verdon}}, \bibinfo{journal}{Physics of Plasmas}
  \textbf{\bibinfo{volume}{3}}, \bibinfo{pages}{2122} (\bibinfo{year}{1996}).

\bibitem[{\citenamefont{Goncharov
  et~al.}(1996{\natexlab{a}})\citenamefont{Goncharov, Betti, McCrory,
  Sorotokin, and Verdon}}]{goncharov1996self1}
\bibinfo{author}{\bibfnamefont{V.~N.} \bibnamefont{Goncharov}},
  \bibinfo{author}{\bibfnamefont{R.}~\bibnamefont{Betti}},
  \bibinfo{author}{\bibfnamefont{R.~L.} \bibnamefont{McCrory}},
  \bibinfo{author}{\bibfnamefont{P.}~\bibnamefont{Sorotokin}},
  \bibnamefont{and} \bibinfo{author}{\bibfnamefont{C.~P.}
  \bibnamefont{Verdon}}, \bibinfo{journal}{Physics of Plasmas}
  \textbf{\bibinfo{volume}{3}}, \bibinfo{pages}{1402}
  (\bibinfo{year}{1996}{\natexlab{a}}).

\bibitem[{\citenamefont{Goncharov
  et~al.}(1996{\natexlab{b}})\citenamefont{Goncharov, Betti, McCrory, and
  Verdon}}]{goncharov1996self2}
\bibinfo{author}{\bibfnamefont{V.~N.} \bibnamefont{Goncharov}},
  \bibinfo{author}{\bibfnamefont{R.}~\bibnamefont{Betti}},
  \bibinfo{author}{\bibfnamefont{R.~L.} \bibnamefont{McCrory}},
  \bibnamefont{and} \bibinfo{author}{\bibfnamefont{C.~P.}
  \bibnamefont{Verdon}}, \bibinfo{journal}{Physics of Plasmas}
  \textbf{\bibinfo{volume}{3}}, \bibinfo{pages}{4665}
  (\bibinfo{year}{1996}{\natexlab{b}}).

\bibitem[{\citenamefont{Ishizaki and
  Nishihara}(1997)}]{ishizaki1997propagation}
\bibinfo{author}{\bibfnamefont{R.}~\bibnamefont{Ishizaki}} \bibnamefont{and}
  \bibinfo{author}{\bibfnamefont{K.}~\bibnamefont{Nishihara}},
  \bibinfo{journal}{Phys. Rev. Lett.} \textbf{\bibinfo{volume}{78}},
  \bibinfo{pages}{1920} (\bibinfo{year}{1997}).

\bibitem[{\citenamefont{Ishizaki and Nishihara}(1998)}]{ishizaki1998model}
\bibinfo{author}{\bibfnamefont{R.}~\bibnamefont{Ishizaki}} \bibnamefont{and}
  \bibinfo{author}{\bibfnamefont{K.}~\bibnamefont{Nishihara}},
  \bibinfo{journal}{Phys. Rev. E} \textbf{\bibinfo{volume}{58}},
  \bibinfo{pages}{3744} (\bibinfo{year}{1998}).

\bibitem[{\citenamefont{Igumenshchev et~al.}(2013)\citenamefont{Igumenshchev,
  Goncharov, Shmayda, Harding, Sangster, and
  Meyerhofer}}]{igumenshchev2013effects}
\bibinfo{author}{\bibfnamefont{I.~V.} \bibnamefont{Igumenshchev}},
  \bibinfo{author}{\bibfnamefont{V.~N.} \bibnamefont{Goncharov}},
  \bibinfo{author}{\bibfnamefont{W.~T.} \bibnamefont{Shmayda}},
  \bibinfo{author}{\bibfnamefont{D.~R.} \bibnamefont{Harding}},
  \bibinfo{author}{\bibfnamefont{T.~C.} \bibnamefont{Sangster}},
  \bibnamefont{and} \bibinfo{author}{\bibfnamefont{D.~D.}
  \bibnamefont{Meyerhofer}}, \bibinfo{journal}{Physics of Plasmas}
  \textbf{\bibinfo{volume}{20}}, \bibinfo{pages}{082703}
  (\bibinfo{year}{2013}).

\bibitem[{\citenamefont{Hu et~al.}(2016)\citenamefont{Hu, Michel, Davis, Betti,
  Radha, Campbell, Froula, and Stoeckl}}]{hu2016understanding}
\bibinfo{author}{\bibfnamefont{S.~X.} \bibnamefont{Hu}},
  \bibinfo{author}{\bibfnamefont{D.~T.} \bibnamefont{Michel}},
  \bibinfo{author}{\bibfnamefont{A.~K.} \bibnamefont{Davis}},
  \bibinfo{author}{\bibfnamefont{R.}~\bibnamefont{Betti}},
  \bibinfo{author}{\bibfnamefont{P.~B.} \bibnamefont{Radha}},
  \bibinfo{author}{\bibfnamefont{E.~M.} \bibnamefont{Campbell}},
  \bibinfo{author}{\bibfnamefont{D.~H.} \bibnamefont{Froula}},
  \bibnamefont{and} \bibinfo{author}{\bibfnamefont{C.}~\bibnamefont{Stoeckl}},
  \bibinfo{journal}{Physics of Plasmas} \textbf{\bibinfo{volume}{23}},
  \bibinfo{pages}{102701} (\bibinfo{year}{2016}).

\bibitem[{\citenamefont{Radha et~al.}(2005)\citenamefont{Radha, Goncharov,
  Collins, Delettrez, Elbaz, Glebov, Keck, Keller, Knauer, Marozas
  et~al.}}]{radha2005two}
\bibinfo{author}{\bibfnamefont{P.~B.} \bibnamefont{Radha}},
  \bibinfo{author}{\bibfnamefont{V.~N.} \bibnamefont{Goncharov}},
  \bibinfo{author}{\bibfnamefont{T.~J.~B.} \bibnamefont{Collins}},
  \bibinfo{author}{\bibfnamefont{J.~A.} \bibnamefont{Delettrez}},
  \bibinfo{author}{\bibfnamefont{Y.}~\bibnamefont{Elbaz}},
  \bibinfo{author}{\bibfnamefont{V.~Y.} \bibnamefont{Glebov}},
  \bibinfo{author}{\bibfnamefont{R.~L.} \bibnamefont{Keck}},
  \bibinfo{author}{\bibfnamefont{D.~E.} \bibnamefont{Keller}},
  \bibinfo{author}{\bibfnamefont{J.~P.} \bibnamefont{Knauer}},
  \bibinfo{author}{\bibfnamefont{J.~A.} \bibnamefont{Marozas}},
  \bibnamefont{et~al.}, \bibinfo{journal}{Physics of Plasmas}
  \textbf{\bibinfo{volume}{12}}, \bibinfo{pages}{032702}
  (\bibinfo{year}{2005}).

\bibitem[{\citenamefont{Hu et~al.}(2009)\citenamefont{Hu, Radha, Marozas,
  Betti, Collins, Craxton, Delettrez, Edgell, Epstein, Goncharov
  et~al.}}]{hu2009neutron}
\bibinfo{author}{\bibfnamefont{S.~X.} \bibnamefont{Hu}},
  \bibinfo{author}{\bibfnamefont{P.~B.} \bibnamefont{Radha}},
  \bibinfo{author}{\bibfnamefont{J.~A.} \bibnamefont{Marozas}},
  \bibinfo{author}{\bibfnamefont{R.}~\bibnamefont{Betti}},
  \bibinfo{author}{\bibfnamefont{T.~J.~B.} \bibnamefont{Collins}},
  \bibinfo{author}{\bibfnamefont{R.~S.} \bibnamefont{Craxton}},
  \bibinfo{author}{\bibfnamefont{J.~A.} \bibnamefont{Delettrez}},
  \bibinfo{author}{\bibfnamefont{D.~H.} \bibnamefont{Edgell}},
  \bibinfo{author}{\bibfnamefont{R.}~\bibnamefont{Epstein}},
  \bibinfo{author}{\bibfnamefont{V.~N.} \bibnamefont{Goncharov}},
  \bibnamefont{et~al.}, \bibinfo{journal}{Physics of Plasmas}
  \textbf{\bibinfo{volume}{16}}, \bibinfo{pages}{112706}
  (\bibinfo{year}{2009}).

\bibitem[{\citenamefont{Hu et~al.}(2010)\citenamefont{Hu, Goncharov, Radha,
  Marozas, Skupsky, Boehly, Sangster, Meyerhofer, and McCrory}}]{hu2010two}
\bibinfo{author}{\bibfnamefont{S.~X.} \bibnamefont{Hu}},
  \bibinfo{author}{\bibfnamefont{V.~N.} \bibnamefont{Goncharov}},
  \bibinfo{author}{\bibfnamefont{P.~B.} \bibnamefont{Radha}},
  \bibinfo{author}{\bibfnamefont{J.~A.} \bibnamefont{Marozas}},
  \bibinfo{author}{\bibfnamefont{S.}~\bibnamefont{Skupsky}},
  \bibinfo{author}{\bibfnamefont{T.~R.} \bibnamefont{Boehly}},
  \bibinfo{author}{\bibfnamefont{T.~C.} \bibnamefont{Sangster}},
  \bibinfo{author}{\bibfnamefont{D.~D.} \bibnamefont{Meyerhofer}},
  \bibnamefont{and} \bibinfo{author}{\bibfnamefont{R.~L.}
  \bibnamefont{McCrory}}, \bibinfo{journal}{Physics of Plasmas}
  \textbf{\bibinfo{volume}{17}}, \bibinfo{pages}{102706}
  (\bibinfo{year}{2010}).

\bibitem[{\citenamefont{Baumgaertel et~al.}(2014)\citenamefont{Baumgaertel,
  Bradley, Hsu, Cobble, Hakel, Tregillis, Krasheninnikova, Murphy, Schmitt,
  Shah et~al.}}]{baumgaertel2014observation}
\bibinfo{author}{\bibfnamefont{J.~A.} \bibnamefont{Baumgaertel}},
  \bibinfo{author}{\bibfnamefont{P.~A.} \bibnamefont{Bradley}},
  \bibinfo{author}{\bibfnamefont{S.~C.} \bibnamefont{Hsu}},
  \bibinfo{author}{\bibfnamefont{J.~A.} \bibnamefont{Cobble}},
  \bibinfo{author}{\bibfnamefont{P.}~\bibnamefont{Hakel}},
  \bibinfo{author}{\bibfnamefont{I.~L.} \bibnamefont{Tregillis}},
  \bibinfo{author}{\bibfnamefont{N.~S.} \bibnamefont{Krasheninnikova}},
  \bibinfo{author}{\bibfnamefont{T.~J.} \bibnamefont{Murphy}},
  \bibinfo{author}{\bibfnamefont{M.~J.} \bibnamefont{Schmitt}},
  \bibinfo{author}{\bibfnamefont{R.~C.} \bibnamefont{Shah}},
  \bibnamefont{et~al.}, \bibinfo{journal}{Physics of Plasmas}
  \textbf{\bibinfo{volume}{21}}, \bibinfo{pages}{052706}
  (\bibinfo{year}{2014}).

\bibitem[{\citenamefont{Metzler et~al.}(2003)\citenamefont{Metzler, Velikovich,
  Schmitt, Karasik, Serlin, Mostovych, Obenschain, Gardner, and
  Aglitskiy}}]{metzler2003laser}
\bibinfo{author}{\bibfnamefont{N.}~\bibnamefont{Metzler}},
  \bibinfo{author}{\bibfnamefont{A.~L.} \bibnamefont{Velikovich}},
  \bibinfo{author}{\bibfnamefont{A.~J.} \bibnamefont{Schmitt}},
  \bibinfo{author}{\bibfnamefont{M.}~\bibnamefont{Karasik}},
  \bibinfo{author}{\bibfnamefont{V.}~\bibnamefont{Serlin}},
  \bibinfo{author}{\bibfnamefont{A.~N.} \bibnamefont{Mostovych}},
  \bibinfo{author}{\bibfnamefont{S.~P.} \bibnamefont{Obenschain}},
  \bibinfo{author}{\bibfnamefont{J.~H.} \bibnamefont{Gardner}},
  \bibnamefont{and}
  \bibinfo{author}{\bibfnamefont{Y.}~\bibnamefont{Aglitskiy}},
  \bibinfo{journal}{Physics of Plasmas} \textbf{\bibinfo{volume}{10}},
  \bibinfo{pages}{1897} (\bibinfo{year}{2003}).

\bibitem[{\citenamefont{Goncharov et~al.}(2003)\citenamefont{Goncharov, Knauer,
  McKenty, Radha, Sangster, Skupsky, Betti, McCrory, and
  Meyerhofer}}]{goncharov2003improved}
\bibinfo{author}{\bibfnamefont{V.~N.} \bibnamefont{Goncharov}},
  \bibinfo{author}{\bibfnamefont{J.~P.} \bibnamefont{Knauer}},
  \bibinfo{author}{\bibfnamefont{P.~W.} \bibnamefont{McKenty}},
  \bibinfo{author}{\bibfnamefont{P.~B.} \bibnamefont{Radha}},
  \bibinfo{author}{\bibfnamefont{T.~C.} \bibnamefont{Sangster}},
  \bibinfo{author}{\bibfnamefont{S.}~\bibnamefont{Skupsky}},
  \bibinfo{author}{\bibfnamefont{R.}~\bibnamefont{Betti}},
  \bibinfo{author}{\bibfnamefont{R.~L.} \bibnamefont{McCrory}},
  \bibnamefont{and} \bibinfo{author}{\bibfnamefont{D.~D.}
  \bibnamefont{Meyerhofer}}, \bibinfo{journal}{Physics of Plasmas}
  \textbf{\bibinfo{volume}{10}}, \bibinfo{pages}{1906} (\bibinfo{year}{2003}).

\bibitem[{\citenamefont{Anderson and Betti}(2004)}]{anderson2004laser}
\bibinfo{author}{\bibfnamefont{K.}~\bibnamefont{Anderson}} \bibnamefont{and}
  \bibinfo{author}{\bibfnamefont{R.}~\bibnamefont{Betti}},
  \bibinfo{journal}{Physics of Plasmas} \textbf{\bibinfo{volume}{11}},
  \bibinfo{pages}{5} (\bibinfo{year}{2004}).

\bibitem[{\citenamefont{Watt et~al.}(1998)\citenamefont{Watt, Duke, Fontes,
  Gobby, Hollis, Kopp, Mason, Wilson, Verdon, Boehly et~al.}}]{watt1998laser}
\bibinfo{author}{\bibfnamefont{R.~G.} \bibnamefont{Watt}},
  \bibinfo{author}{\bibfnamefont{J.}~\bibnamefont{Duke}},
  \bibinfo{author}{\bibfnamefont{C.~J.} \bibnamefont{Fontes}},
  \bibinfo{author}{\bibfnamefont{P.~L.} \bibnamefont{Gobby}},
  \bibinfo{author}{\bibfnamefont{R.~V.} \bibnamefont{Hollis}},
  \bibinfo{author}{\bibfnamefont{R.~A.} \bibnamefont{Kopp}},
  \bibinfo{author}{\bibfnamefont{R.~J.} \bibnamefont{Mason}},
  \bibinfo{author}{\bibfnamefont{D.~C.} \bibnamefont{Wilson}},
  \bibinfo{author}{\bibfnamefont{C.~P.} \bibnamefont{Verdon}},
  \bibinfo{author}{\bibfnamefont{T.~R.} \bibnamefont{Boehly}},
  \bibnamefont{et~al.}, \bibinfo{journal}{Phys. Rev. Lett.}
  \textbf{\bibinfo{volume}{81}}, \bibinfo{pages}{4644} (\bibinfo{year}{1998}).

\bibitem[{\citenamefont{Metzler et~al.}(2002)\citenamefont{Metzler, Velikovich,
  Schmitt, and Gardner}}]{metzler2002laser}
\bibinfo{author}{\bibfnamefont{N.}~\bibnamefont{Metzler}},
  \bibinfo{author}{\bibfnamefont{A.~L.} \bibnamefont{Velikovich}},
  \bibinfo{author}{\bibfnamefont{A.~J.} \bibnamefont{Schmitt}},
  \bibnamefont{and} \bibinfo{author}{\bibfnamefont{J.~H.}
  \bibnamefont{Gardner}}, \bibinfo{journal}{Physics of Plasmas}
  \textbf{\bibinfo{volume}{9}}, \bibinfo{pages}{5050} (\bibinfo{year}{2002}).

\bibitem[{\citenamefont{Hu et~al.}(2018)\citenamefont{Hu, Theobald, Radha,
  Peebles, Regan, Nikroo, Bonino, Harding, Goncharov, Petta
  et~al.}}]{hu2018mitigating}
\bibinfo{author}{\bibfnamefont{S.~X.} \bibnamefont{Hu}},
  \bibinfo{author}{\bibfnamefont{W.}~\bibnamefont{Theobald}},
  \bibinfo{author}{\bibfnamefont{P.~B.} \bibnamefont{Radha}},
  \bibinfo{author}{\bibfnamefont{J.~L.} \bibnamefont{Peebles}},
  \bibinfo{author}{\bibfnamefont{S.~P.} \bibnamefont{Regan}},
  \bibinfo{author}{\bibfnamefont{A.}~\bibnamefont{Nikroo}},
  \bibinfo{author}{\bibfnamefont{M.~J.} \bibnamefont{Bonino}},
  \bibinfo{author}{\bibfnamefont{D.~R.} \bibnamefont{Harding}},
  \bibinfo{author}{\bibfnamefont{V.~N.} \bibnamefont{Goncharov}},
  \bibinfo{author}{\bibfnamefont{N.}~\bibnamefont{Petta}},
  \bibnamefont{et~al.}, \bibinfo{journal}{Physics of Plasmas}
  \textbf{\bibinfo{volume}{25}}, \bibinfo{pages}{082710}
  (\bibinfo{year}{2018}).

\bibitem[{\citenamefont{Masse}(2007)}]{masse2007stabilizing}
\bibinfo{author}{\bibfnamefont{L.}~\bibnamefont{Masse}},
  \bibinfo{journal}{Phys. Rev. Lett.} \textbf{\bibinfo{volume}{98}},
  \bibinfo{pages}{245001} (\bibinfo{year}{2007}).

\bibitem[{\citenamefont{Masse et~al.}(2011)\citenamefont{Masse, Casner,
  Galmiche, Huser, Liberatore, and Theobald}}]{masse2011observation}
\bibinfo{author}{\bibfnamefont{L.}~\bibnamefont{Masse}},
  \bibinfo{author}{\bibfnamefont{A.}~\bibnamefont{Casner}},
  \bibinfo{author}{\bibfnamefont{D.}~\bibnamefont{Galmiche}},
  \bibinfo{author}{\bibfnamefont{G.}~\bibnamefont{Huser}},
  \bibinfo{author}{\bibfnamefont{S.}~\bibnamefont{Liberatore}},
  \bibnamefont{and} \bibinfo{author}{\bibfnamefont{M.}~\bibnamefont{Theobald}},
  \bibinfo{journal}{Phys. Rev. E} \textbf{\bibinfo{volume}{83}},
  \bibinfo{pages}{055401} (\bibinfo{year}{2011}).

\bibitem[{\citenamefont{Hu et~al.}(2012)\citenamefont{Hu, Fiksel, Goncharov,
  Skupsky, Meyerhofer, and Smalyuk}}]{hu2012mitigating}
\bibinfo{author}{\bibfnamefont{S.~X.} \bibnamefont{Hu}},
  \bibinfo{author}{\bibfnamefont{G.}~\bibnamefont{Fiksel}},
  \bibinfo{author}{\bibfnamefont{V.~N.} \bibnamefont{Goncharov}},
  \bibinfo{author}{\bibfnamefont{S.}~\bibnamefont{Skupsky}},
  \bibinfo{author}{\bibfnamefont{D.~D.} \bibnamefont{Meyerhofer}},
  \bibnamefont{and} \bibinfo{author}{\bibfnamefont{V.~A.}
  \bibnamefont{Smalyuk}}, \bibinfo{journal}{Phys. Rev. Lett.}
  \textbf{\bibinfo{volume}{108}}, \bibinfo{pages}{195003}
  (\bibinfo{year}{2012}).

\bibitem[{\citenamefont{Fiksel et~al.}(2012)\citenamefont{Fiksel, Hu,
  Goncharov, Meyerhofer, Sangster, Smalyuk, Yaakobi, Bonino, and
  Jungquist}}]{fiksel2012experimental}
\bibinfo{author}{\bibfnamefont{G.}~\bibnamefont{Fiksel}},
  \bibinfo{author}{\bibfnamefont{S.~X.} \bibnamefont{Hu}},
  \bibinfo{author}{\bibfnamefont{V.~A.} \bibnamefont{Goncharov}},
  \bibinfo{author}{\bibfnamefont{D.~D.} \bibnamefont{Meyerhofer}},
  \bibinfo{author}{\bibfnamefont{T.~C.} \bibnamefont{Sangster}},
  \bibinfo{author}{\bibfnamefont{V.~A.} \bibnamefont{Smalyuk}},
  \bibinfo{author}{\bibfnamefont{B.}~\bibnamefont{Yaakobi}},
  \bibinfo{author}{\bibfnamefont{M.~J.} \bibnamefont{Bonino}},
  \bibnamefont{and}
  \bibinfo{author}{\bibfnamefont{R.}~\bibnamefont{Jungquist}},
  \bibinfo{journal}{Physics of Plasmas} \textbf{\bibinfo{volume}{19}},
  \bibinfo{pages}{062704} (\bibinfo{year}{2012}).

\bibitem[{\citenamefont{Obenschain et~al.}(2002)\citenamefont{Obenschain,
  Colombant, Karasik, Pawley, Serlin, Schmitt, Weaver, Gardner, Phillips,
  Aglitskiy et~al.}}]{obenschain2002effects}
\bibinfo{author}{\bibfnamefont{S.~P.} \bibnamefont{Obenschain}},
  \bibinfo{author}{\bibfnamefont{D.~G.} \bibnamefont{Colombant}},
  \bibinfo{author}{\bibfnamefont{M.}~\bibnamefont{Karasik}},
  \bibinfo{author}{\bibfnamefont{C.~J.} \bibnamefont{Pawley}},
  \bibinfo{author}{\bibfnamefont{V.}~\bibnamefont{Serlin}},
  \bibinfo{author}{\bibfnamefont{A.~J.} \bibnamefont{Schmitt}},
  \bibinfo{author}{\bibfnamefont{J.~L.} \bibnamefont{Weaver}},
  \bibinfo{author}{\bibfnamefont{J.~H.} \bibnamefont{Gardner}},
  \bibinfo{author}{\bibfnamefont{L.}~\bibnamefont{Phillips}},
  \bibinfo{author}{\bibfnamefont{Y.}~\bibnamefont{Aglitskiy}},
  \bibnamefont{et~al.}, \bibinfo{journal}{Physics of Plasmas}
  \textbf{\bibinfo{volume}{9}}, \bibinfo{pages}{2234} (\bibinfo{year}{2002}).

\bibitem[{\citenamefont{Mostovych et~al.}(2008)\citenamefont{Mostovych,
  Colombant, Karasik, Knauer, Schmitt, and Weaver}}]{mostovych2008enhanced}
\bibinfo{author}{\bibfnamefont{A.~N.} \bibnamefont{Mostovych}},
  \bibinfo{author}{\bibfnamefont{D.~G.} \bibnamefont{Colombant}},
  \bibinfo{author}{\bibfnamefont{M.}~\bibnamefont{Karasik}},
  \bibinfo{author}{\bibfnamefont{J.~P.} \bibnamefont{Knauer}},
  \bibinfo{author}{\bibfnamefont{A.~J.} \bibnamefont{Schmitt}},
  \bibnamefont{and} \bibinfo{author}{\bibfnamefont{J.~L.}
  \bibnamefont{Weaver}}, \bibinfo{journal}{Phys. Rev. Lett.}
  \textbf{\bibinfo{volume}{100}}, \bibinfo{pages}{075002}
  (\bibinfo{year}{2008}).

\bibitem[{\citenamefont{Karasik et~al.}(2015)\citenamefont{Karasik, Weaver,
  Aglitskiy, Oh, and Obenschain}}]{karasik2015suppression}
\bibinfo{author}{\bibfnamefont{M.}~\bibnamefont{Karasik}},
  \bibinfo{author}{\bibfnamefont{J.~L.} \bibnamefont{Weaver}},
  \bibinfo{author}{\bibfnamefont{Y.}~\bibnamefont{Aglitskiy}},
  \bibinfo{author}{\bibfnamefont{J.}~\bibnamefont{Oh}}, \bibnamefont{and}
  \bibinfo{author}{\bibfnamefont{S.~P.} \bibnamefont{Obenschain}},
  \bibinfo{journal}{Phys. Rev. Lett.} \textbf{\bibinfo{volume}{114}},
  \bibinfo{pages}{085001} (\bibinfo{year}{2015}).

\bibitem[{\citenamefont{Ramis et~al.}(1988)\citenamefont{Ramis, Schmalz, and
  Meyer-Ter-Vehn}}]{ramis1988multi}
\bibinfo{author}{\bibfnamefont{R.}~\bibnamefont{Ramis}},
  \bibinfo{author}{\bibfnamefont{R.}~\bibnamefont{Schmalz}}, \bibnamefont{and}
  \bibinfo{author}{\bibfnamefont{J.}~\bibnamefont{Meyer-Ter-Vehn}},
  \bibinfo{journal}{Computer Physics Communications}
  \textbf{\bibinfo{volume}{49}}, \bibinfo{pages}{475} (\bibinfo{year}{1988}),
  ISSN \bibinfo{issn}{0010-4655}.

\bibitem[{\citenamefont{Ramis et~al.}(2012)\citenamefont{Ramis, Eidmann, ter
  Vehn, and Hüller}}]{ramis2012multi}
\bibinfo{author}{\bibfnamefont{R.}~\bibnamefont{Ramis}},
  \bibinfo{author}{\bibfnamefont{K.}~\bibnamefont{Eidmann}},
  \bibinfo{author}{\bibfnamefont{J.~M.} \bibnamefont{ter Vehn}},
  \bibnamefont{and} \bibinfo{author}{\bibfnamefont{S.}~\bibnamefont{Hüller}},
  \bibinfo{journal}{Computer Physics Communications}
  \textbf{\bibinfo{volume}{183}}, \bibinfo{pages}{637} (\bibinfo{year}{2012}),
  ISSN \bibinfo{issn}{0010-4655}.

\bibitem[{\citenamefont{Ramis and ter Vehn}(2016)}]{ramis2016multi}
\bibinfo{author}{\bibfnamefont{R.}~\bibnamefont{Ramis}} \bibnamefont{and}
  \bibinfo{author}{\bibfnamefont{J.~M.} \bibnamefont{ter Vehn}},
  \bibinfo{journal}{Computer Physics Communications}
  \textbf{\bibinfo{volume}{203}}, \bibinfo{pages}{226} (\bibinfo{year}{2016}),
  ISSN \bibinfo{issn}{0010-4655}.

\bibitem[{\citenamefont{Poli et~al.}(2007)\citenamefont{Poli, Kennedy, and
  Blackwell}}]{poli2007particle}
\bibinfo{author}{\bibfnamefont{R.}~\bibnamefont{Poli}},
  \bibinfo{author}{\bibfnamefont{J.}~\bibnamefont{Kennedy}}, \bibnamefont{and}
  \bibinfo{author}{\bibfnamefont{T.}~\bibnamefont{Blackwell}},
  \bibinfo{journal}{Swarm Intelligence} \textbf{\bibinfo{volume}{1}},
  \bibinfo{pages}{33} (\bibinfo{year}{2007}), ISSN \bibinfo{issn}{1935-3820}.

\bibitem[{\citenamefont{Fryxell et~al.}(2000)\citenamefont{Fryxell, Olson,
  Ricker, Timmes, Zingale, Lamb, MacNeice, Rosner, Truran, and
  Tufo}}]{Fryxell_2000}
\bibinfo{author}{\bibfnamefont{B.}~\bibnamefont{Fryxell}},
  \bibinfo{author}{\bibfnamefont{K.}~\bibnamefont{Olson}},
  \bibinfo{author}{\bibfnamefont{P.}~\bibnamefont{Ricker}},
  \bibinfo{author}{\bibfnamefont{F.~X.} \bibnamefont{Timmes}},
  \bibinfo{author}{\bibfnamefont{M.}~\bibnamefont{Zingale}},
  \bibinfo{author}{\bibfnamefont{D.~Q.} \bibnamefont{Lamb}},
  \bibinfo{author}{\bibfnamefont{P.}~\bibnamefont{MacNeice}},
  \bibinfo{author}{\bibfnamefont{R.}~\bibnamefont{Rosner}},
  \bibinfo{author}{\bibfnamefont{J.~W.} \bibnamefont{Truran}},
  \bibnamefont{and} \bibinfo{author}{\bibfnamefont{H.}~\bibnamefont{Tufo}},
  \bibinfo{journal}{The Astrophysical Journal Supplement Series}
  \textbf{\bibinfo{volume}{131}}, \bibinfo{pages}{273} (\bibinfo{year}{2000}).

\bibitem[{\citenamefont{Zhang et~al.}(2020)\citenamefont{Zhang, Wang, Yang, Wu,
  Ma, Jiao, Zhang, Wu, Yuan, Li et~al.}}]{zhang2020double}
\bibinfo{author}{\bibfnamefont{J.}~\bibnamefont{Zhang}},
  \bibinfo{author}{\bibfnamefont{W.~M.} \bibnamefont{Wang}},
  \bibinfo{author}{\bibfnamefont{X.~H.} \bibnamefont{Yang}},
  \bibinfo{author}{\bibfnamefont{D.}~\bibnamefont{Wu}},
  \bibinfo{author}{\bibfnamefont{Y.~Y.} \bibnamefont{Ma}},
  \bibinfo{author}{\bibfnamefont{J.~L.} \bibnamefont{Jiao}},
  \bibinfo{author}{\bibfnamefont{Z.}~\bibnamefont{Zhang}},
  \bibinfo{author}{\bibfnamefont{F.~Y.} \bibnamefont{Wu}},
  \bibinfo{author}{\bibfnamefont{X.~H.} \bibnamefont{Yuan}},
  \bibinfo{author}{\bibfnamefont{Y.~T.} \bibnamefont{Li}},
  \bibnamefont{et~al.}, \bibinfo{journal}{Philosophical Transactions of the
  Royal Society A: Mathematical, Physical and Engineering Sciences}
  \textbf{\bibinfo{volume}{378}}, \bibinfo{pages}{20200015}
  (\bibinfo{year}{2020}).

\bibitem[{\citenamefont{Abdallah~Jr and Clark}(1985)}]{abdallah1985tops}
\bibinfo{author}{\bibfnamefont{J.}~\bibnamefont{Abdallah~Jr}} \bibnamefont{and}
  \bibinfo{author}{\bibfnamefont{R.~E.} \bibnamefont{Clark}},
  \bibinfo{type}{Tech. Rep.}, \bibinfo{institution}{Los Alamos National Lab.,
  NM (USA)} (\bibinfo{year}{1985}).

\bibitem[{\citenamefont{{Magee} et~al.}(1995)\citenamefont{{Magee}, {Abdallah},
  {Clark}, {Cohen}, {Collins}, {Csanak}, {Fontes}, {Gauger}, {Keady},
  {Kilcrease} et~al.}}]{magee1995atomic}
\bibinfo{author}{\bibfnamefont{N.~H.} \bibnamefont{{Magee}}},
  \bibinfo{author}{\bibfnamefont{J.}~\bibnamefont{{Abdallah}},
  \bibfnamefont{J.}}, \bibinfo{author}{\bibfnamefont{R.~E.~H.}
  \bibnamefont{{Clark}}}, \bibinfo{author}{\bibfnamefont{J.~S.}
  \bibnamefont{{Cohen}}}, \bibinfo{author}{\bibfnamefont{L.~A.}
  \bibnamefont{{Collins}}},
  \bibinfo{author}{\bibfnamefont{G.}~\bibnamefont{{Csanak}}},
  \bibinfo{author}{\bibfnamefont{C.~J.} \bibnamefont{{Fontes}}},
  \bibinfo{author}{\bibfnamefont{A.}~\bibnamefont{{Gauger}}},
  \bibinfo{author}{\bibfnamefont{J.~J.} \bibnamefont{{Keady}}},
  \bibinfo{author}{\bibfnamefont{D.~P.} \bibnamefont{{Kilcrease}}},
  \bibnamefont{et~al.}, in \emph{\bibinfo{booktitle}{Astrophysical Applications
  of Powerful New Databases}}, edited by \bibinfo{editor}{\bibfnamefont{S.~J.}
  \bibnamefont{{Adelman}}} \bibnamefont{and}
  \bibinfo{editor}{\bibfnamefont{W.~L.} \bibnamefont{{Wiese}}}
  (\bibinfo{year}{1995}), vol.~\bibinfo{volume}{78} of
  \emph{\bibinfo{series}{Astronomical Society of the Pacific Conference
  Series}}, p.~\bibinfo{pages}{51}.

\bibitem[{\citenamefont{Zel'dovich and Raizer}(2002)}]{zel2002physics}
\bibinfo{author}{\bibfnamefont{Y.}~\bibnamefont{Zel'dovich}} \bibnamefont{and}
  \bibinfo{author}{\bibfnamefont{Y.}~\bibnamefont{Raizer}},
  \emph{\bibinfo{title}{Physics of Shock Waves and High-Temperature
  Hydrodynamic Phenomena}}, Dover Books on Physics (\bibinfo{publisher}{Dover
  Publications}, \bibinfo{year}{2002}), ISBN \bibinfo{isbn}{9780486420028}.

\bibitem[{\citenamefont{Atzeni and Meyer-ter Vehn}(2009)}]{atzeni2004physics}
\bibinfo{author}{\bibfnamefont{S.}~\bibnamefont{Atzeni}} \bibnamefont{and}
  \bibinfo{author}{\bibfnamefont{J.}~\bibnamefont{Meyer-ter Vehn}},
  \emph{\bibinfo{title}{The Physics of Inertial Fusion: Beam Plasma
  Interaction, Hydrodynamics, Hot Dense Matter}}, International Series of
  Monographs on Physics (\bibinfo{publisher}{Oxford University Press},
  \bibinfo{year}{2009}), ISBN \bibinfo{isbn}{9780199568017}.

\bibitem[{\citenamefont{Betti et~al.}(1998)\citenamefont{Betti, Goncharov,
  McCrory, and Verdon}}]{betti1998growth}
\bibinfo{author}{\bibfnamefont{R.}~\bibnamefont{Betti}},
  \bibinfo{author}{\bibfnamefont{V.~N.} \bibnamefont{Goncharov}},
  \bibinfo{author}{\bibfnamefont{R.~L.} \bibnamefont{McCrory}},
  \bibnamefont{and} \bibinfo{author}{\bibfnamefont{C.~P.}
  \bibnamefont{Verdon}}, \bibinfo{journal}{Physics of Plasmas}
  \textbf{\bibinfo{volume}{5}}, \bibinfo{pages}{1446} (\bibinfo{year}{1998}).

\end{thebibliography}
	
\end{document}